\documentclass[10pt,letterpaper]{article}
\usepackage{opex3}

\begin{document}

\title{Enhanced transmission through subwavelength apertures by excitation
of particle localized plasmons and nanojets}
\author{F.J. Valdivia-Valero$^{1}$ and M. Nieto-Vesperinas$^{1,*}$}
\address{$^{1}$Instituto de Ciencia de Materiales de Madrid, C.S.I.C., Campus de Cantoblanco \\ 28049 Madrid, Spain}
\email{$^{*}$mnieto@icmm.csic.es}


\begin{abstract}
We study, and illustrate with numerical calculations,  transmission
enhancement by subwavelength 2D slits due to the dominant role
played by the excitation of the eigenmodes of {\it plasmonic}
cylinders when they are placed at the aperture entrance; and also
due to reinforced and highly localized energy in the slit as a
consequence of the formation of a {\it nanojet}. We show that,
providing the illumination is chosen such that an aperture
transmitting eigenmode is generated, the phenomenon is independent
of whether or not the slit alone produces extraordinary
transmission; although in the former case this enhancement will add
to this slit supertransmission. We address several particle sizes,
and emphasize the universality of this phenomenon with different
materials.
\end{abstract}

\ocis{(050.1940) Diffraction; (050.1220) Apertures; (050.6624)
Subwavelength structures; (160.4236) Nanomaterials; (230.5750)
Resonators; (230.7370) Waveguides; (240.6680) Surface plasmons;
(250.6715) Switching.}


\section{Introduction}




Extraordinary transmission of light through apertures of size
smaller than the wavelength $\lambda$ has been a subject with a vast
amount of  interest \cite{GarciadeAbajo2007, MartinMoreno2010} since
the first experiments \cite{Ebbesen1998} demonstrated that light
could pass through subwavelength hole arrays of radius $r$, as well
as through single apertures with a transmitted intensity much larger
than as predicted by Bethe's theory \cite{Bethe1944}:
$(r/\lambda)^4$. Many subsequent works \cite{MartinMoreno2001,
Lezec2002, GarciadeAbajo2002, Barnes2004, Gordon2004, Degiron2004,
GarciaVidal2005, Webb2006} deepened both theoretically
\cite{MartinMoreno2001} and experimentally \cite{Degiron2004,
Lezec2004} in the understanding of this phenomenon for apertures
practiced either in metallic or even in dielectric slabs; and
proposals were made to even increase this transmission by a single
aperture, for instance by introducing periodic corrugation on the
slab around a single hole \cite{Lezec2002}, such that the beaming
effect of the zero order of this grating could produce more
directionally transmitted light through the hole. Other research to
increase the efficiency of this process included employing other
different aperture geometries \cite{Degiron2004, GarciaVidal2005} or
placing metamaterial slabs in front of the aperture \cite{Alu2006}.
Much work has been undertaken at optical frequencies, but there are
also investigations at THz \cite{GomezRivas2003} as well as using
the resonances of split ring resonators SRRs in the microwave
regimes \cite{Ozbay2009_1, Ozbay2009_2}. In particular with the SRRs
they experimentally showed \cite{Ozbay2009_2} 740-fold transmission
enhancement by amplifying the incoming wave into the subwavelength
aperture with the help of the SRRs. Subsequent studies
\cite{Soukulis2010} even increased this transmittance amount.

The underlying concept for all these configurations is to produce a
resonance that gives rise to intense and highly localized fields
close to the aperture entrance, so that they couple with the
resonance of the aperture, thus leading to enhanced fields at its
exit and therefore to an enhanced transmission.

In this paper we study and illustrate this mechanism with numerical
calculations capable of predicting or reproducing experimental
results with great accuracy, but this time we shall employ as the
resonant trigger a particle Mie resonance, in this case a {\it
plasmonic nanoparticle} close to the subwavelength aperture
entrance. Then the localized intense fields at the aperture will be
due to those of the localized surface plasmon mode $LSP_{mn}$ (where
$m$ and $n$ stand for the eigenmode radial and angular numbers) of
the particle which will subsequently couple and enhance the aperture
mode $TE_{uv}$ in its cavity, (where $u$ and $v$ stand for the mode
numbers) and thus its transmittivity, hereafter referred to as an
observed supertransmission peak. Plasmonic particles have not yet
been employed for this task as far as we know.

In addition, since such a large intensity distribution, highly
localized in the aperture, gives rise to an enhanced transmission,
we inquire on the comparison of this plasmon morphology dependent
resonance (MDR) coupling mechanism, which is resonant, to another
one which is not; i. e. that of highly localized light focalization
at the aperture by {\it excitation of a nanojet} \cite{Taflove2004,
Taflove2005, Taflove2006, Taflove2009}. This last effect also gives
rise to aperture transmittance enhancement, as we will show.
Differences between those two effects as regards both the
transmission peak and the transmittance spectrum shape, are
discussed.

\section{Numerical calculations}

We shall employ a finite element method with FEMLAB of COMSOL
(\mbox{\url{http://www.comsol.com}}) and will address a 2D
configuration, since the essential features observed on coupling and
resonance excitations are likewise obtained in 3D \cite{Taflove2004,
Ho1994}. Hence, we simulate the whole physical process through these
numerical calculations. Details of the meshing geometry and
convergence which leads to accurate results are given elsewhere
\cite{GarciaPomar-NietoVesperinas2005,
ValdiviaValero-NietoVesperinas2010}. In the 2D geometry dealt with
here, we have employed an incident wave, linearly p - polarized,
namely with its magnetic vector ${\bf H}_{z}$ perpendicular to the
geometry of the XY - plane (i. e. the plane of the images shown in
this work), and propagating in the Y - axis direction. In the case
of plasmon excitation such a wave has a Gaussian profile at its
focus: ${\bf H_{z}(x, y)}= {\bf
|H_{z0}|}exp(-x^{2}/2\sigma^{2})exp(i((2\pi/\lambda)y-\omega t))$,
$|{\bf H_{z0}}|$ being the modulus amplitude, $2^{1/2}\sigma$ being
the half width at half maximum (HWHM) of the beam, and $\lambda$ its
wavelength. In the case of nanojet excitation, a rectangular profile
corresponding to a tapered time harmonic plane wave ${\bf H_{z}(x,
y)}= {\bf |H_{z0}|}exp(i((2\pi/\lambda)y-\omega t))$ has been
employed. In both cases, wave profiles have been normalized to
unity: ${\bf |H_{z0}|}= 1A/m$ (SI units), this corresponding to an
incident energy flow magnitude $|<{\bf S_{0}}>|\approx 190W/m^{2}$
(SI units). In this way, both {\it plasmonic} and {\it nanojet}
forming particles are cylinders with OZ axis (in fact, nanojets were
first predicted in 2D for infinitely long cylinders
\cite{Taflove2004} and later studied for other 3D particles
\cite{Taflove2005, Taflove2006, Taflove2009}). For such a 2D
geometry, it is well known that this choice of polarization (in
contrast with s - polarization) is the one under which the
subwavelength slit presents homogeneous eigenmodes $TE_{10}$, i. e.
which transmit and may lead to extraordinary transmission
\cite{ValdiviaValero-NietoVesperinas2010, Jackson, Porto1999,
Garcia-NietoVesperinas2007, ValdiviaValero-NietoVesperinas2011}. All
refractive indices of the materials here considered are taken from
\cite{Palik} at the different addressed wavelengths.

\section{Extraordinary transmission enhancement by localized plasmon excitation}

\begin{figure}[htbp]
\begin{minipage}{.49\linewidth}
\centering
\includegraphics[width=7cm]{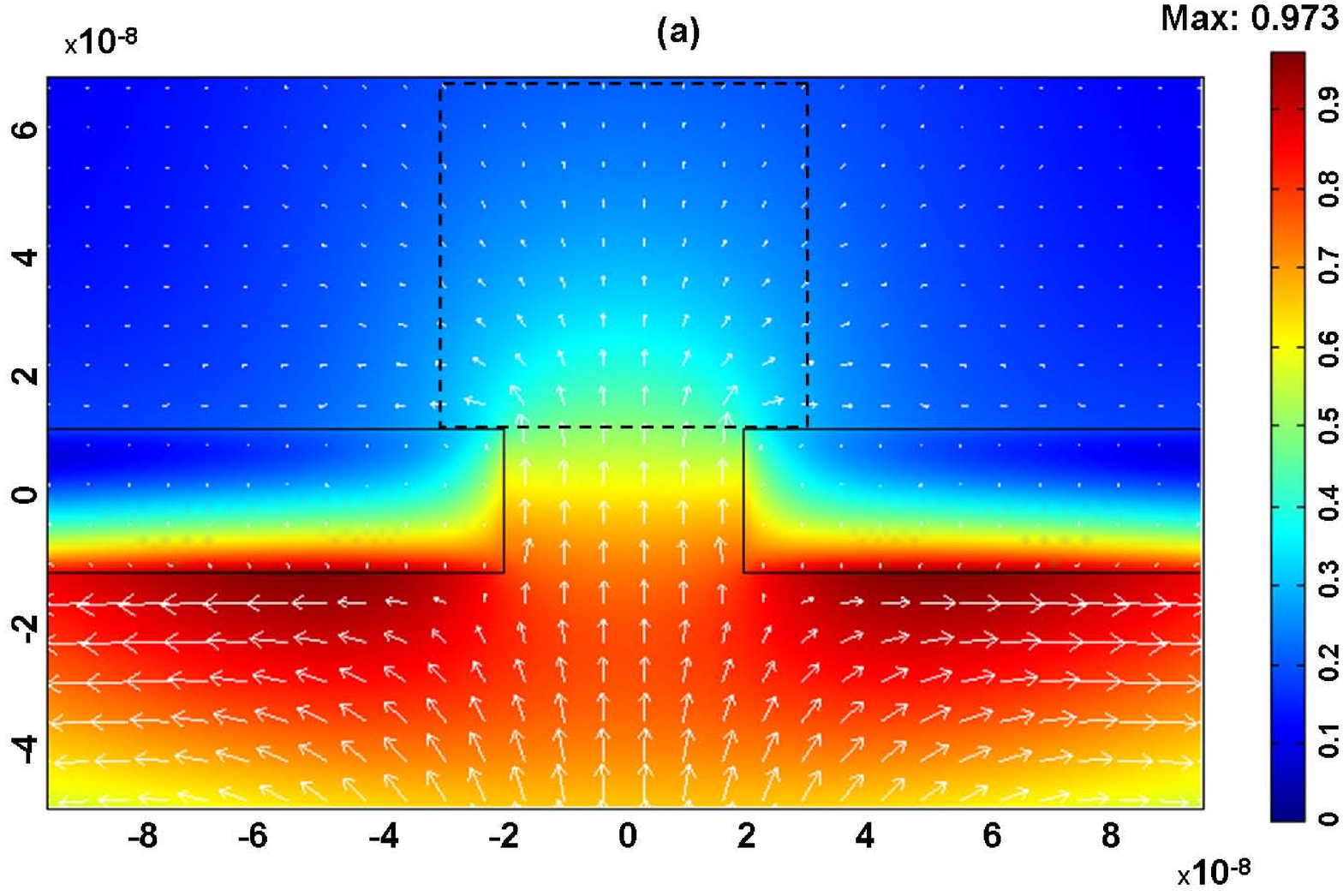}
\end{minipage}
\begin{minipage}{.49\linewidth}
\centering
\includegraphics[width=7cm]{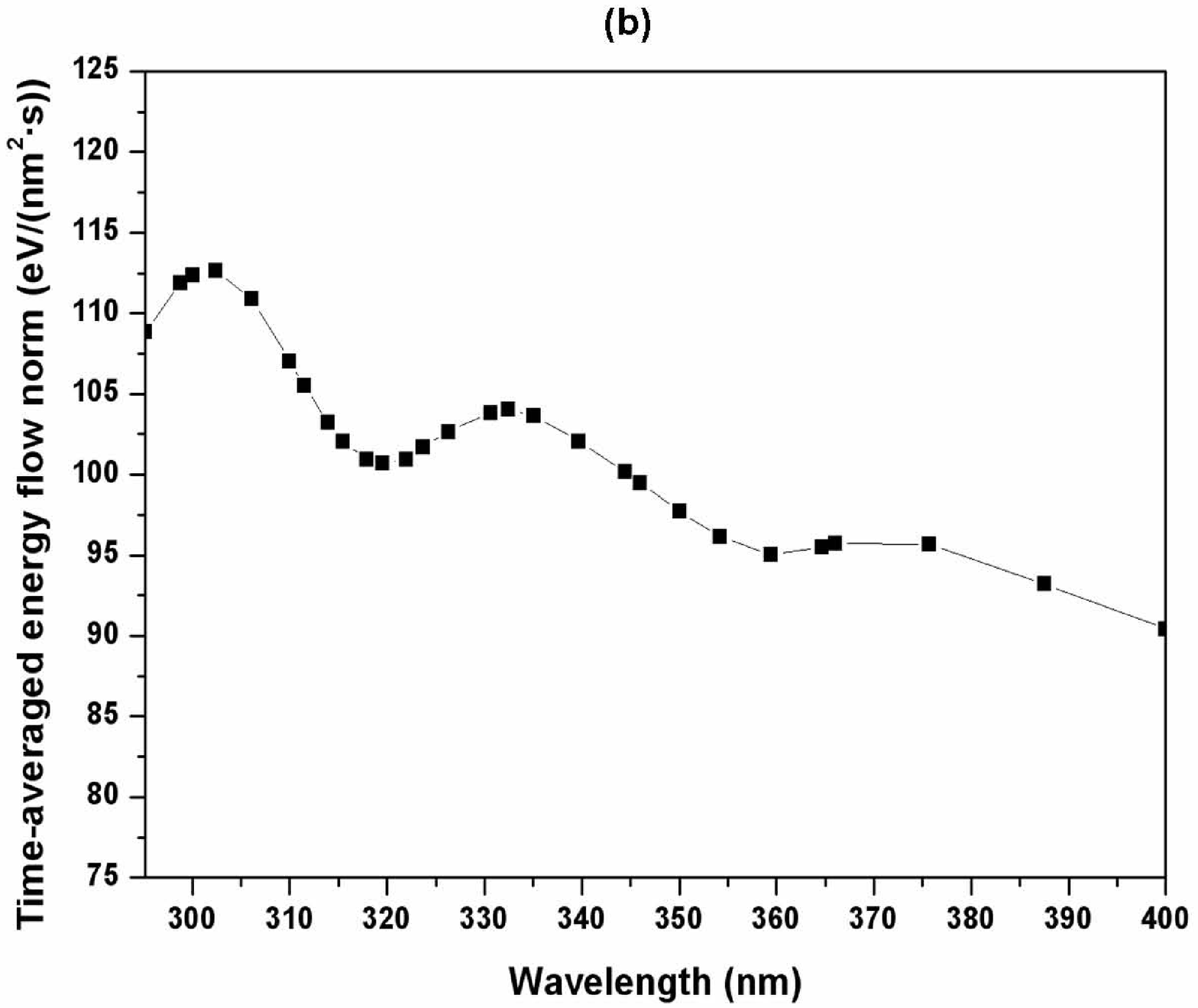}
\end{minipage}
\begin{minipage}{.98\linewidth}
\centering
\includegraphics[width=7cm]{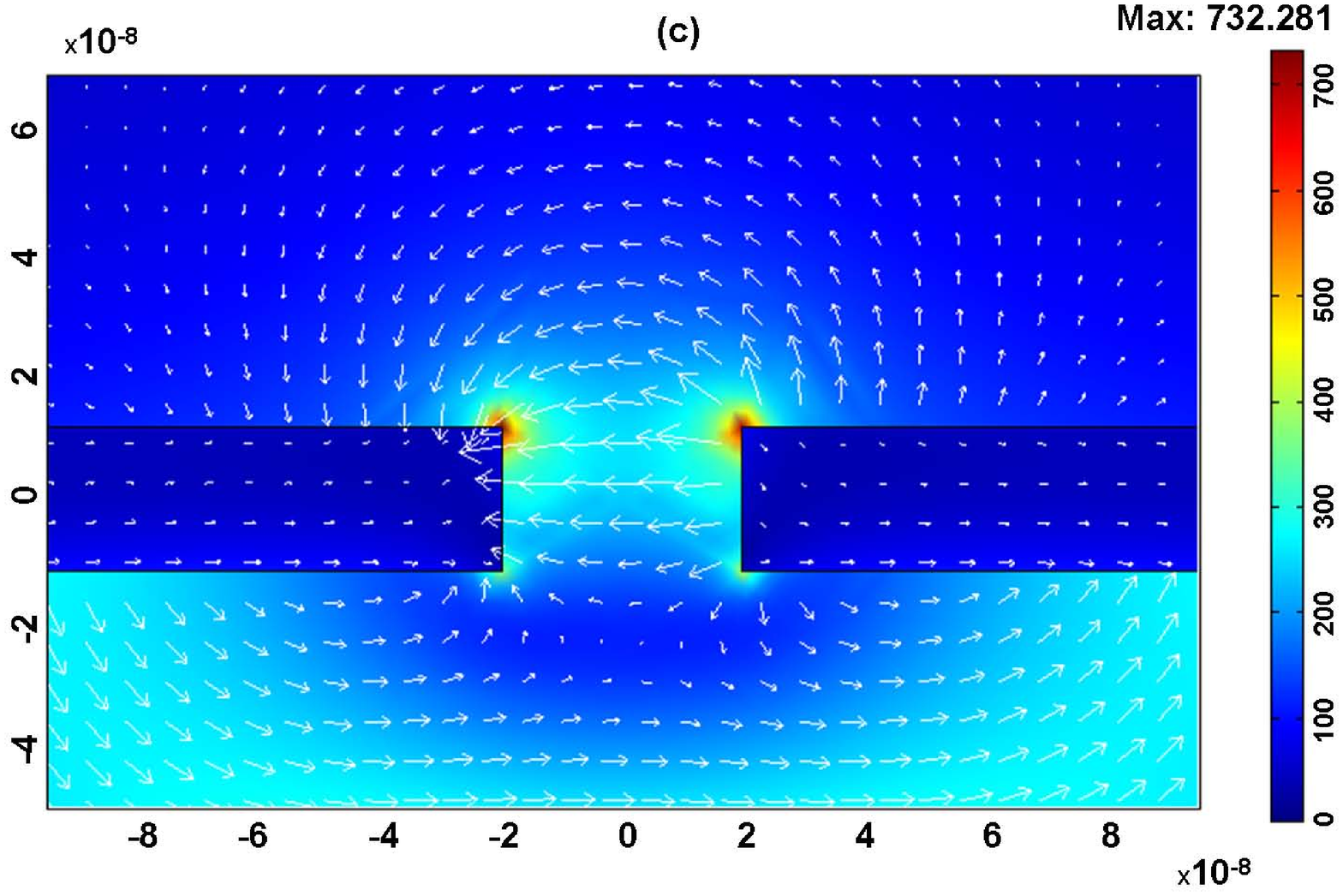}
\end{minipage}
\caption{ (a) $Al$ slab (refractive index $n_{Al}= 0.280 + i3.64$;
slab width $D= 2.61\mu m$; slab thickness $h= 23.76nm$; slit width
$d= 39.59nm$) illuminated by a p - polarized Gaussian beam (${\bf
|H_{z0}|} = 1A/m$; spot size $2^{1/2}\sigma= 55.99nm$) at $\lambda=
302.4nm$ (i. e. exciting the $TE_{10}$ slit mode). The light is
incident upwards from below the slab. Magnetic field norm ${\bf
|H_{z}(r)|}$ (in $A/m$ (SI units)), in colors, and time - averaged
energy flow $<{\bf S(r)}>$, in arrows (maximum arrow length $=
998.41eV/(nm^{2}\cdot s)$). (b) Slab transmission evaluated as
$|<{\bf S(r)}>|$, averaged over a square monitor area $A=
((3/2)d)^{2}= 3526.58nm^{2}$ placed at the exit of the slit (see
Fig. 1(a)). In this range of wavelengths, the highest transmission
peak appears at $\lambda= 302.4nm$. (c) Detail of the electric near
field ${\bf E(r)}$ (in $V/m$ (SI units)) spatial distribution in
both norm (colors) and vector (arrows) for the same configuration as
in Fig. 1(a) at $\lambda= 302.4nm$.}
\end{figure}

Like in some other reported extraordinary transmission experiments
\cite{Ebbesen2005_1, Ebbesen2005_2, Ebbesen2008}, we consider a
metallic {\it Al} slab of thickness $h= 23.76nm$ with a slit of
width $d= 39.59nm$. As mentioned before, the {\it Al} refractive
index at the wavelengths considered here are taken from Ref.
\cite{Palik}, (it should be stressed that re-adapting the
corresponding parameters of the configuration, one could equally
employ other frequently used materials for the slab like, e.g. {\it
Au}. We shall discuss this point in Section 5). The total width of
the slab for the numerical calculation is $D= 2.61\mu m$. This slit
alone is first illuminated with a Gaussian beam and the transmitted
intensity is evaluated at the other side of the slab, close to the
exit of the aperture, in a square of area $A= ((3/2)d)^{2}$, as
shown in Fig. 1(a). In this range of wavelengths, at $\lambda=302.4
nm$ there is a maximum of the light energy transmitted by the slab
into that square monitor, as shown by Fig. 1(b) for different
wavelengths. Figure 1(a) also displays a picture, corresponding to
this wavelength $\lambda= 302.4nm$ of peak transmission, of the
spatial distribution of both ${\bf |H_{z}(r)}|$ and the energy flow
given by the time - averaged Poynting vector ${\bf <S(r)>}$. Whereas
Fig. 1(b) shows the magnitude ${\bf |<S(r)>|}$ transmitted into the
aforementioned square area A at the aperture exit, at different
$\lambda$. On the other hand, Fig. 1(c) illustrates the spatial
distribution of the corresponding electric field near the slab and
aperture. Surface waves, induced on scattering by the aperture
edges, that propagate both along the slab and aperture surfaces are
clearly seen in Fig. 1(a). Also charge concentrations, specially at
the exit corners of the aperture, as a result of the excitation of
the $TE_{10}$ mode inside the aperture, are shown in Figs. 1(a) and
1(c). Specifically, Fig. 1(c) shows that as far as the electric
field contribution is concerned, the induced currents in the
aperture walls associated to the homogeneous $TE_{10}$ mode produce
strong charge concentrations at the vertices of the aperture exit.
These two points of the 2D diagram behave as a dipole giving rise to
a strong electric near field.

\begin{figure}[htbp]
\begin{minipage}{.49\linewidth}
\centering
\includegraphics[width=7cm]{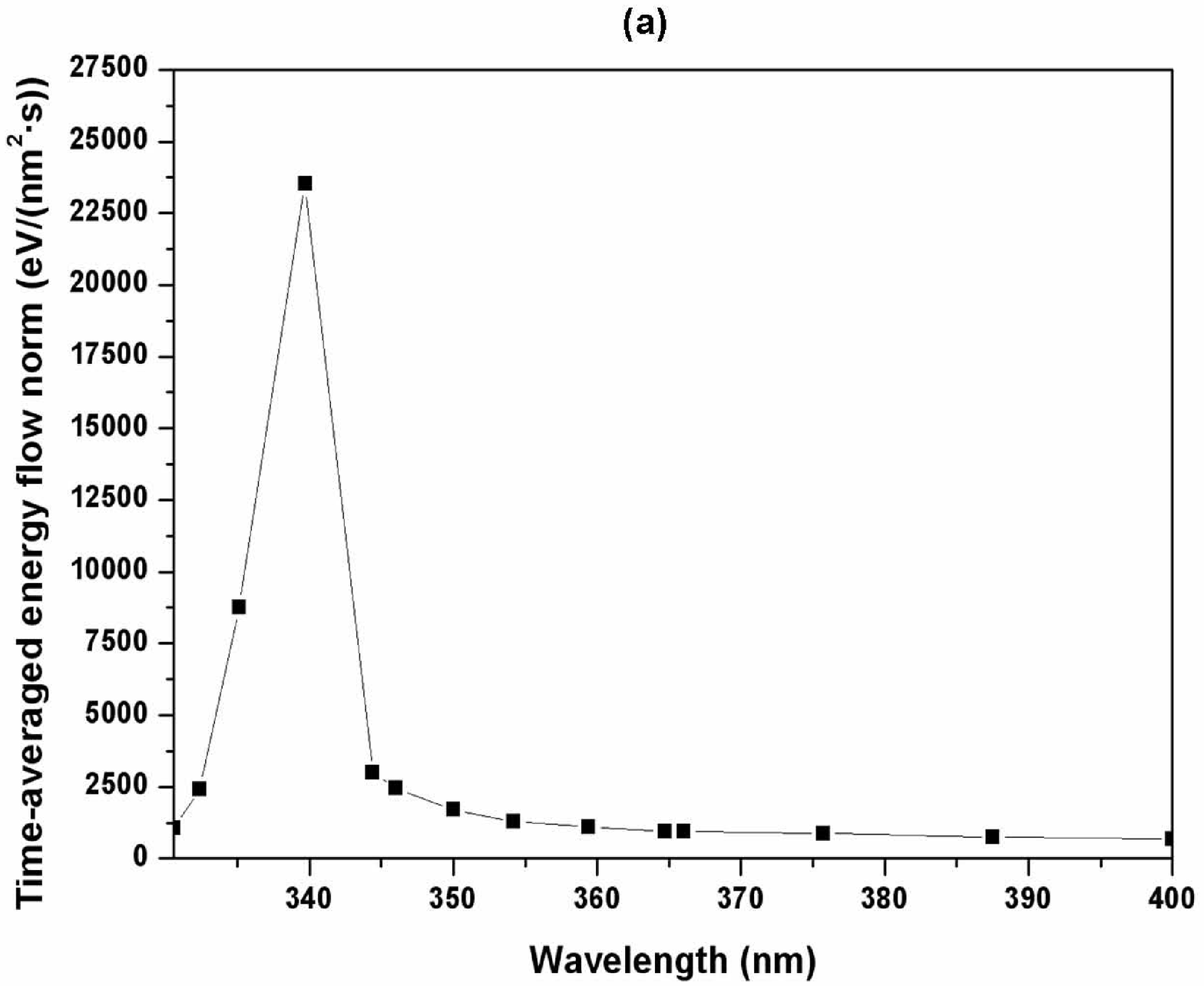}
\end{minipage}
\begin{minipage}{.49\linewidth}
\centering
\includegraphics[width=7cm]{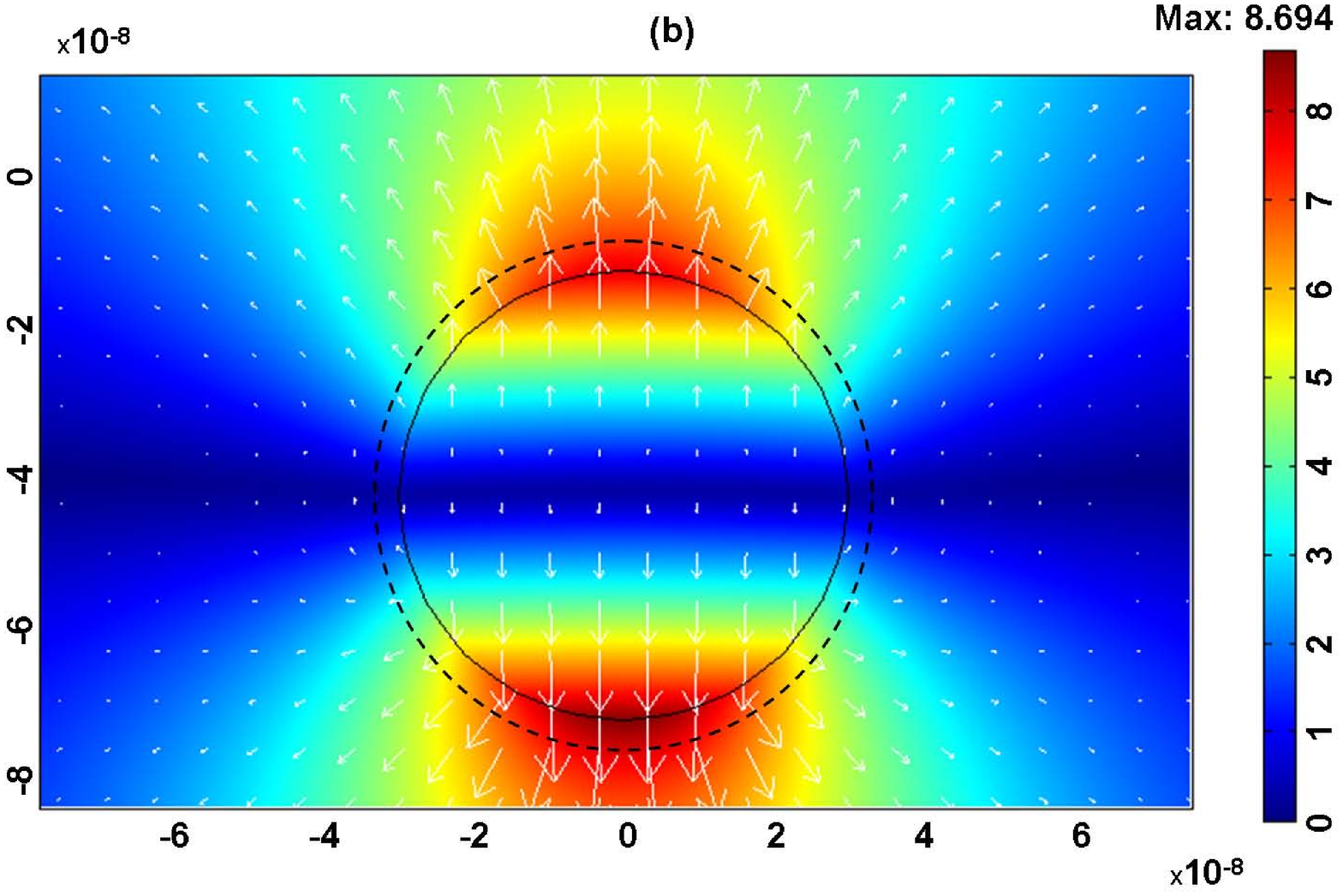}
\end{minipage}
\begin{minipage}{.98\linewidth}
\centering
\includegraphics[width=7cm]{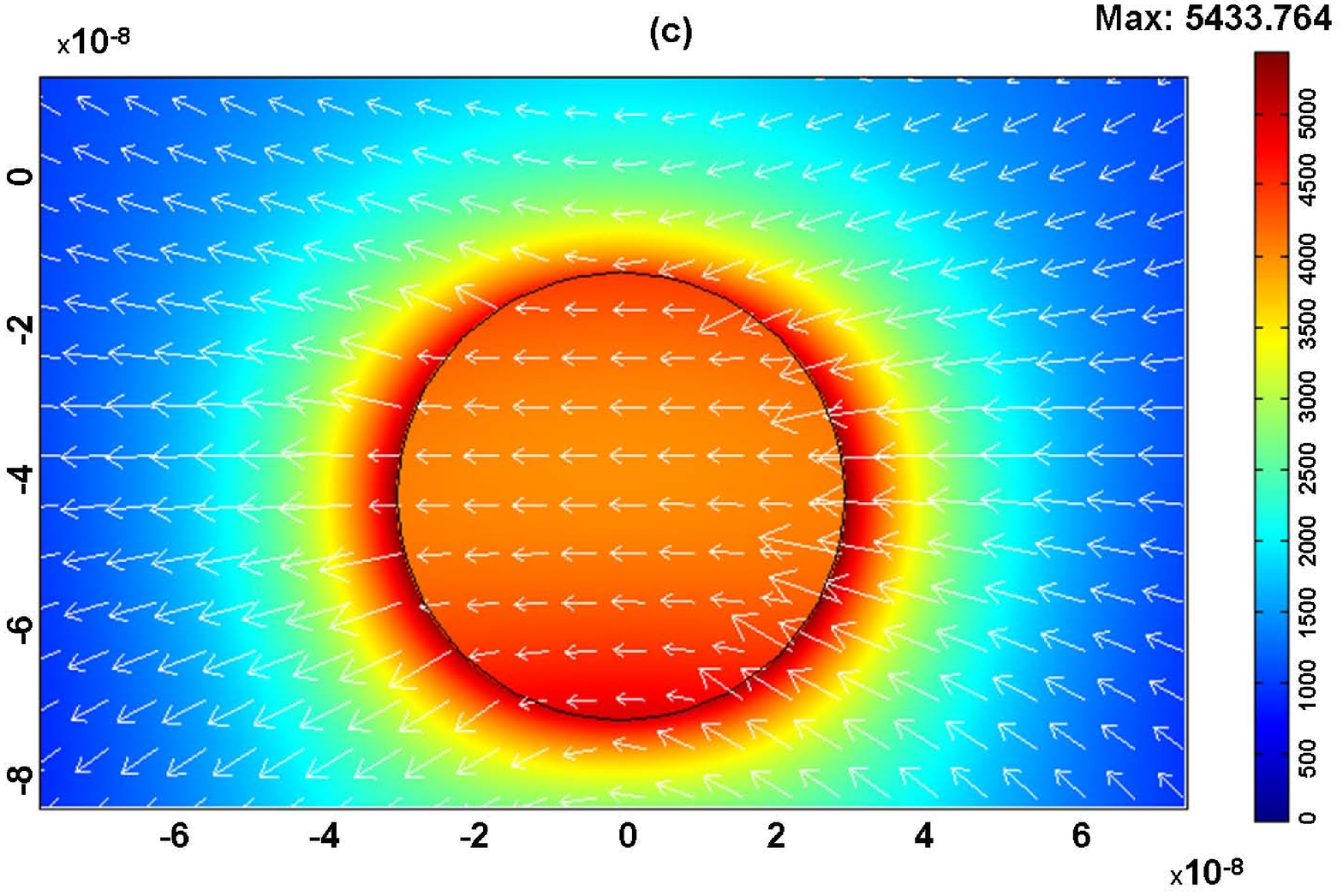}
\end{minipage}
\caption{(a) Response vs. wavelength of an isolated {\it Ag}
cylinder (refractive index $n_{Ag}= 0.259 + i1.12$; radius $r=
30nm$) to a p - polarized Gaussian beam of the same $\sigma$ as in
incident from below with ${\bf |H_{z0}|} = 1A/m$ and $2^{1/2}\sigma
= 59.99nm$, in time - averaged energy flow $|<{\bf S(r)}>|$,
averaged in the area $A= \pi(33^{2}-30^{2})nm^{2}= 593.76nm^{2}$ of
an annulus surrounding the particle (see the two concentric circles
of Fig. 2(b)). The resonance peak occurs at $\lambda= 339.7nm$
(corresponding to the $LSP_{21}$ cylinder mode). (b) {\it Ag}
cylinder of Fig. 2(a), illuminated at $\lambda= 339.7nm$. Spatial
distribution of magnetic field norm $|{\bf H_{z}(r)}|$ (in $A/m$ (SI
units)) in colors, and time - averaged energy flow $<{\bf S(r)}>$ in
arrows (maximum arrow length $= 59071.36eV/(nm^{2}\cdot s)$). (c)
Detail of the electric near field ${\bf E(r)}$ (in $V/m$ (SI units))
distribution in both norm (colors) and vector (arrows) at $\lambda=
339.7nm$.}
\end{figure}

If now a plasmonic cylinder is placed in front of the entrance of
the slit, i.e., at the side of the slab from which the beam is
incident, the transmitted energy by this ensemble: slit plus
particle, may increase dramatically.

To see this, and in order to argument it, let us first consider an
isolated plasmonic cylinder with a Mie resonance at a wavelength not
extremely far from that of the transmission peak of the aperture
alone shown in Figs. 1(a) - 1(c). We shall address an {\it Ag}
\cite{Palik} cylinder of radius $r= 30nm$. The plasmon energy
resonance peak of this cylinder alone, illuminated by the
aforementioned Gaussian beam, occurs at $\lambda=339.7nm$. This is
illustrated by the spectrum of the mean Poynting vector magnitude
near the cylinder in Fig. 2(a), which is evaluated at a narrow
annulus around this particle surface as shown in Fig. 2(b), and
exhibits in both Fig. 2(b) and Fig. 2(c) the characteristic strong
dipolar electromagnetic field intensity distribution and large
concentration of energy flow at this resonant wavelength. A
comparison of the peaks in Fig. 2(a) and Fig. 1(b) already shows
that the magnitude of the energy flow excited at the plasmon
resonance in the {\it Ag} cylinder aforementioned surrounding
annulus, is more than 100 times larger than that transmitted into
the above mentioned exit square monitor by the aperture alone.
Analogous conclusions by a factor 10 are derived for both the
magnetic $|{\bf H_{z}(r)}|$ and electric $|{\bf E(r)}|$ field
modulus which correspond to the isolated aperture and to the
cylinder alone. Also, the excitation wavelengths corresponding to
the peak transmission of the slit and of the cylinder plasmon are
different [cf. Fig. 1(b) and Fig. 2(a)].

\begin{figure}[htbp]
\begin{minipage}{.49\linewidth}
\centering
\includegraphics[width=7cm]{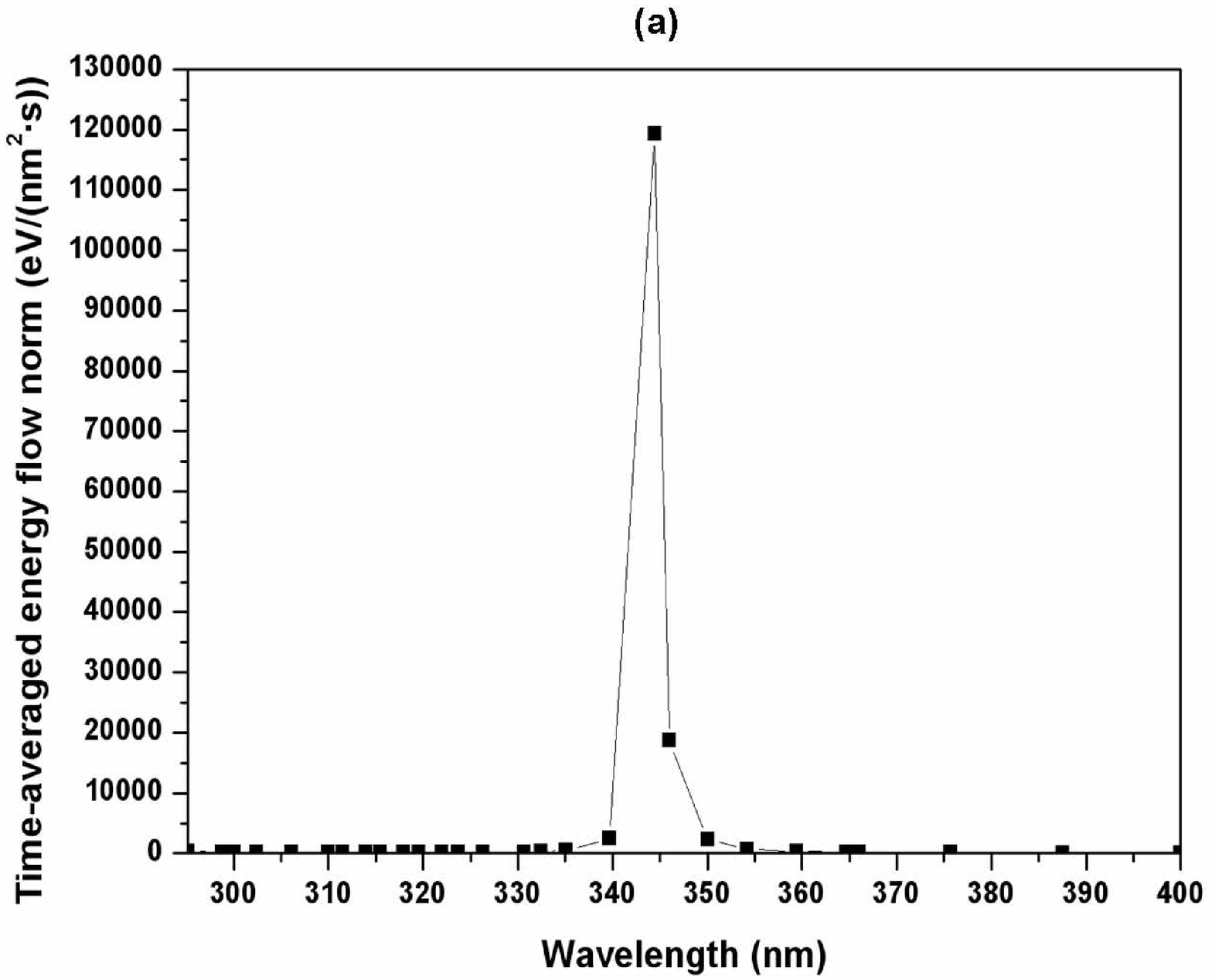}
\end{minipage}
\begin{minipage}{.49\linewidth}
\centering
\includegraphics[width=7cm]{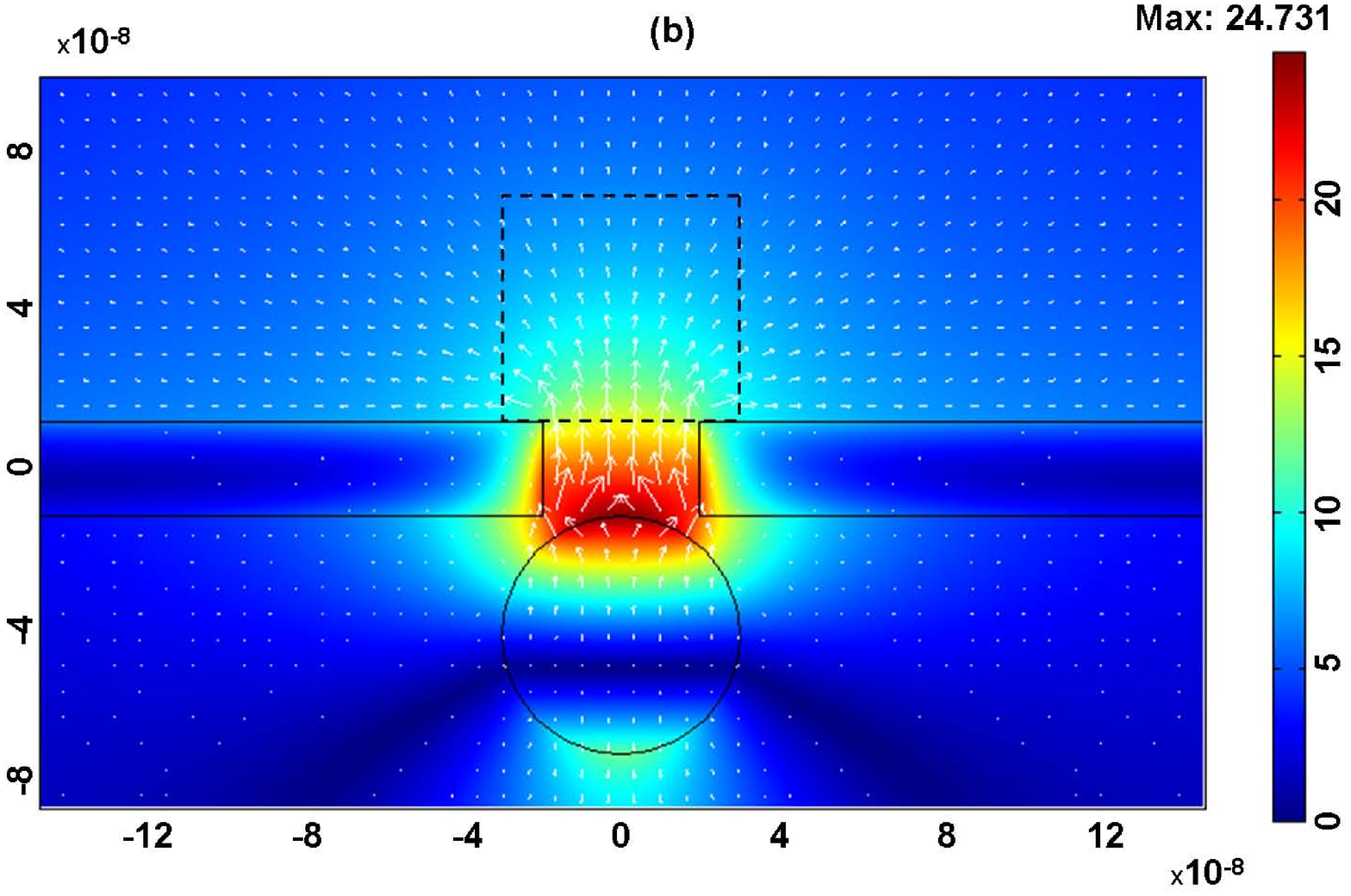}
\end{minipage}
\begin{minipage}{.98\linewidth}
\centering
\includegraphics[width=7cm]{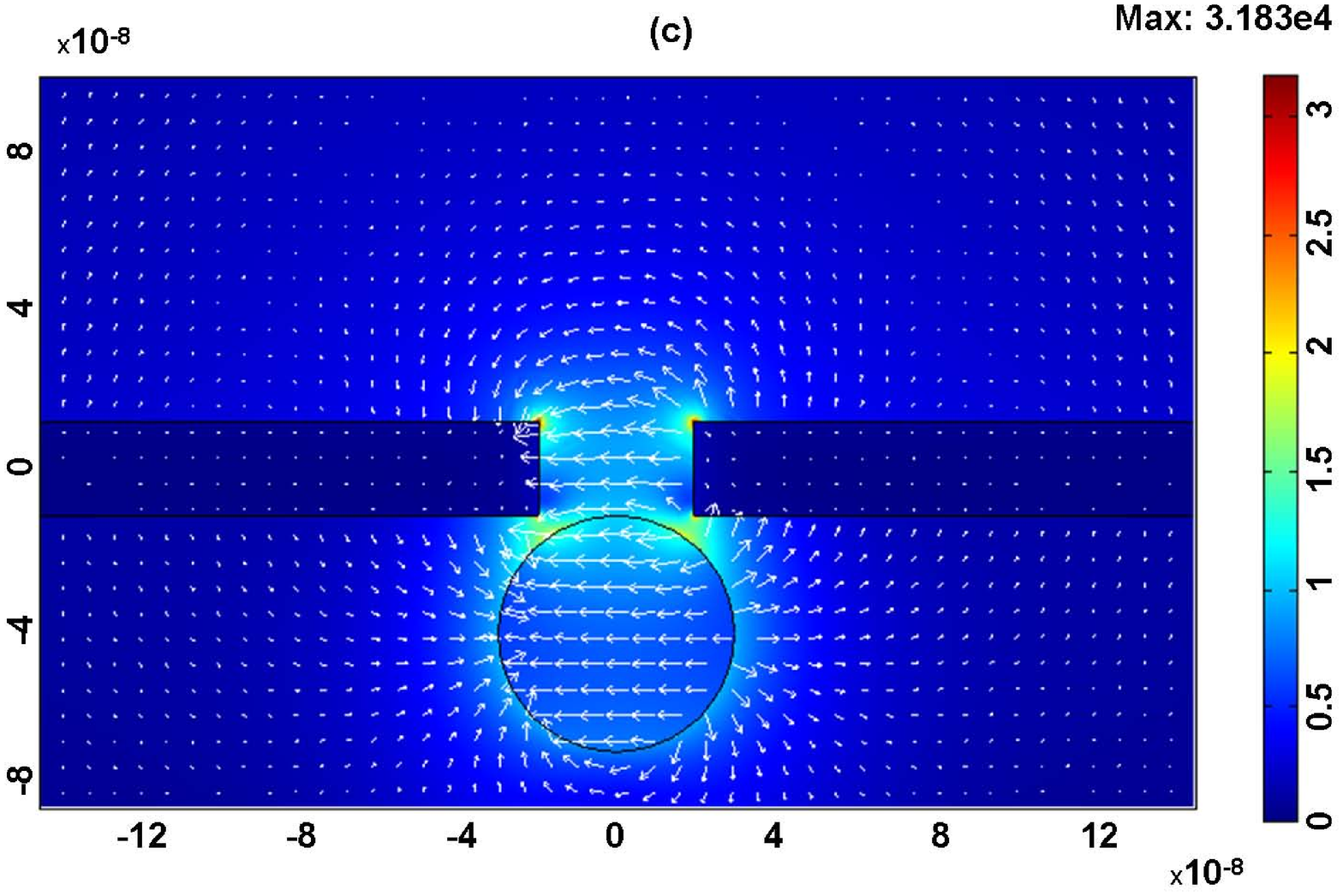}
\end{minipage}
\caption{(a) Transmission of the slit of Fig. 1(a) in presence of
the plasmonic cylinder of Fig. 2(b), placed below the entrance of
this slit as shown in Fig. 3(b). Time - averaged energy flow norm
$|<{\bf S(r)}>|$, averaged over the area of the square monitor shown
in Fig. 3(b), (which is the same as in Fig. 1(a) - 1(b)). The p -
polarized light beam with the same $\sigma$ as in Figs. 1, incides
upwards from below the slab. The highest peak transmission appears
at $\lambda= 344.4nm$ (corresponding to a red-shifted $LSP_{21}$
cylinder resonance). (b) The {\it Ag} cylinder and aperture
illuminated at $\lambda= 344.4nm$ (refractive index $n_{Al}= 0.364 +
i4.17$; $n_{Ag}= 0.238 + i1.24$; showing the magnetic field norm
${\bf |H_{z}(r)|}$ (in $A/m$ (SI units)) in colors, and time -
averaged energy flow $<{\bf S(r)}>$ in arrows (maximum arrow length
$= 946829.51eV/(nm^{2}\cdot s)$). (c) Detail of the electric near
field ${\bf E(r)}$ (in $V/m$ (SI units)) distribution in both norm
(colors) and vector (arrows) for the same configuration as in Fig.
3(b).}
\end{figure}

If we now place this cylinder close to the entrance of the slit as
shown in Figs. 3(a) - 3(c), then the strong energy in this zone, due
to the particle plasmon, will couple and transmit through the
aperture via a feedback particle - slab/aperture, namely both
aperture and particle modes will couple to each other creating a
combined aperture - particle cavity mode at a red-shifted wavelength
with respect to that of the particle alone. Notice that now,
however, the effect of the particle is dominant upon the aperture
transmission. Specifically, the presence of the slab slightly
red-shifts the Mie resonance of the isolated cylinder to
$\lambda=344.4nm$, as expected, but the resulting enhancement of
slit transmission at this new wavelength is now quite large. This is
illustrated in Fig. 3(a) which shows an enhancement of transmitted
energy almost five times larger than that around the particle alone,
also with a much narrower bandwidth and hence with a larger quality
factor, and more than 1000 times that of the transmittance of the
slit alone (cf. Figs. 1(b) and 3(a)). The distributions of magnetic
field and energy flow, as well as of electric field near the
aperture are shown in Fig. 3(b) and Fig. 3(c) respectively, at this
resonant wavelength of $\lambda= 344.4nm$. These showing large
concentrations of field energy in the aperture and especially of
charge and electric field at its exit corners, as well as partial
penetration of the magnetic field through the {\it Al} boundaries of
the slit. This is a phenomenon which usually occurs at thin
apertures at the edges of the aperture. As a matter of fact, we
stress here that {\it this enhancement of the aperture transmission
due to the particle resonance occurs whether or not the aperture
alone does supertransmit or not (of course, as long as the
appropriate polarization, in this case p - waves, is chosen)}. If
the aperture alone already produces extraordinary transmission, the
presence of the resonant particle enhances it.

\begin{figure}[htbp]
\begin{minipage}{.49\linewidth}
\centering
\includegraphics[width=7cm]{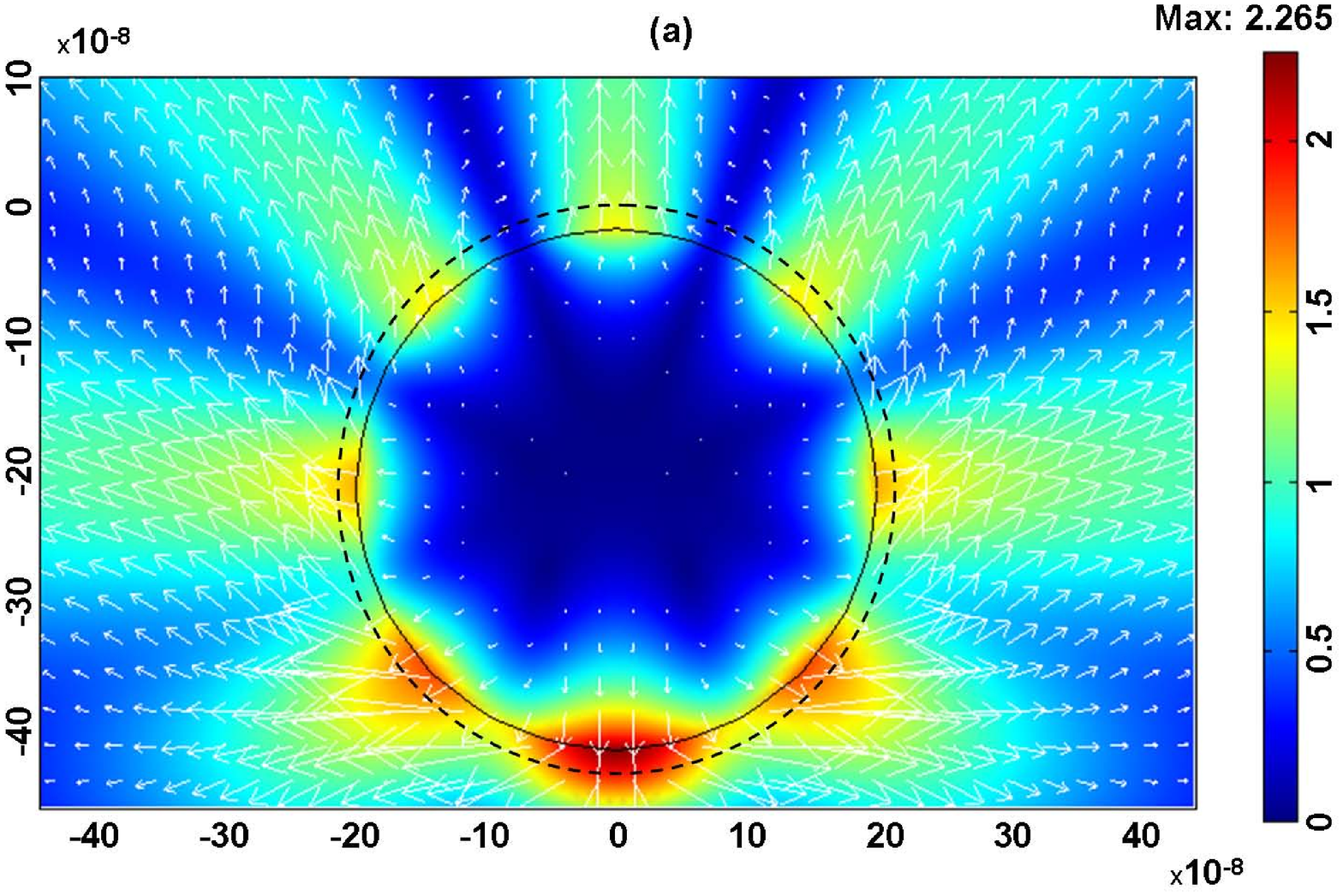}
\end{minipage}
\begin{minipage}{.49\linewidth}
\centering
\includegraphics[width=7cm]{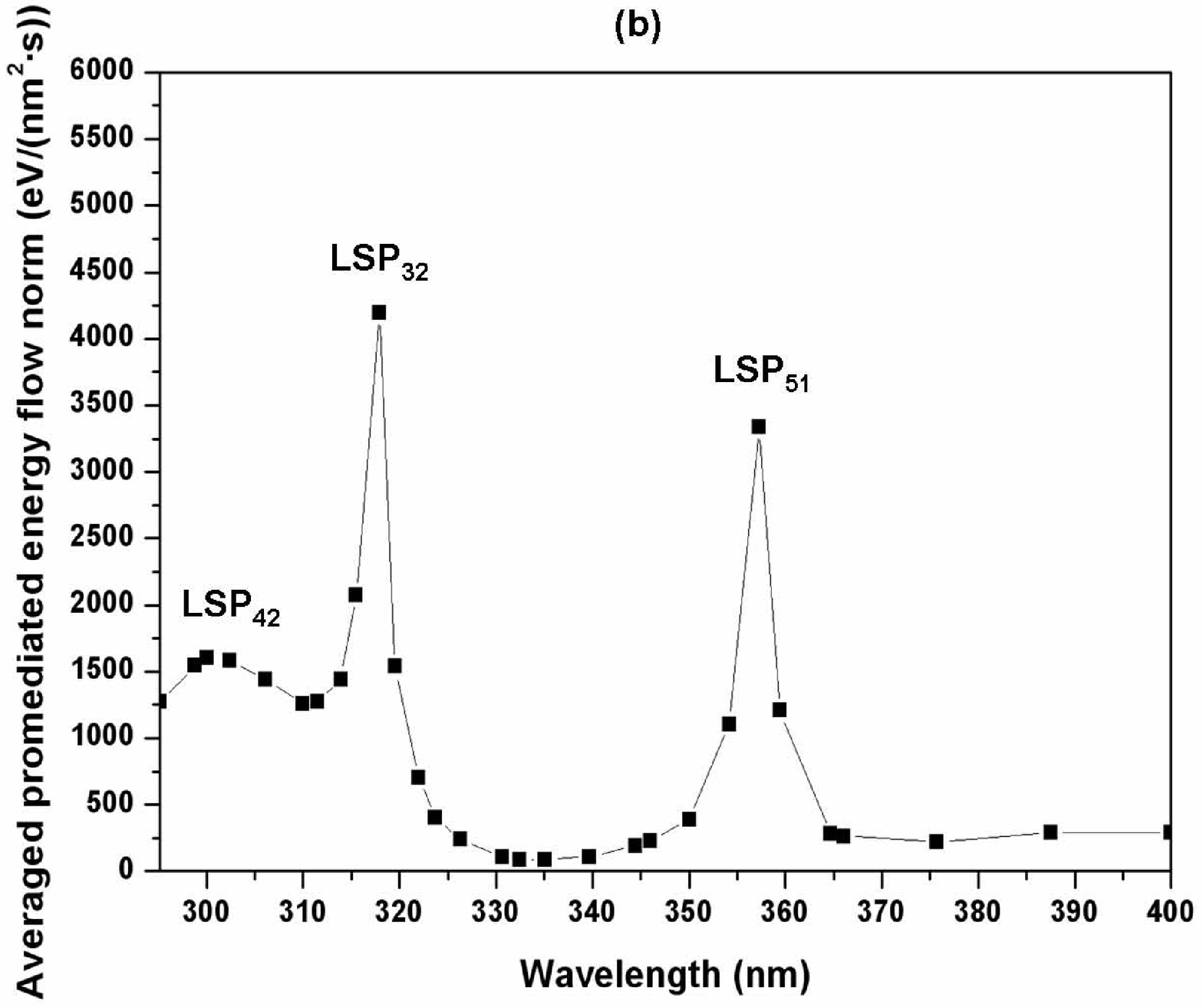}
\end{minipage}
\begin{minipage}{.98\linewidth}
\centering
\includegraphics[width=7cm]{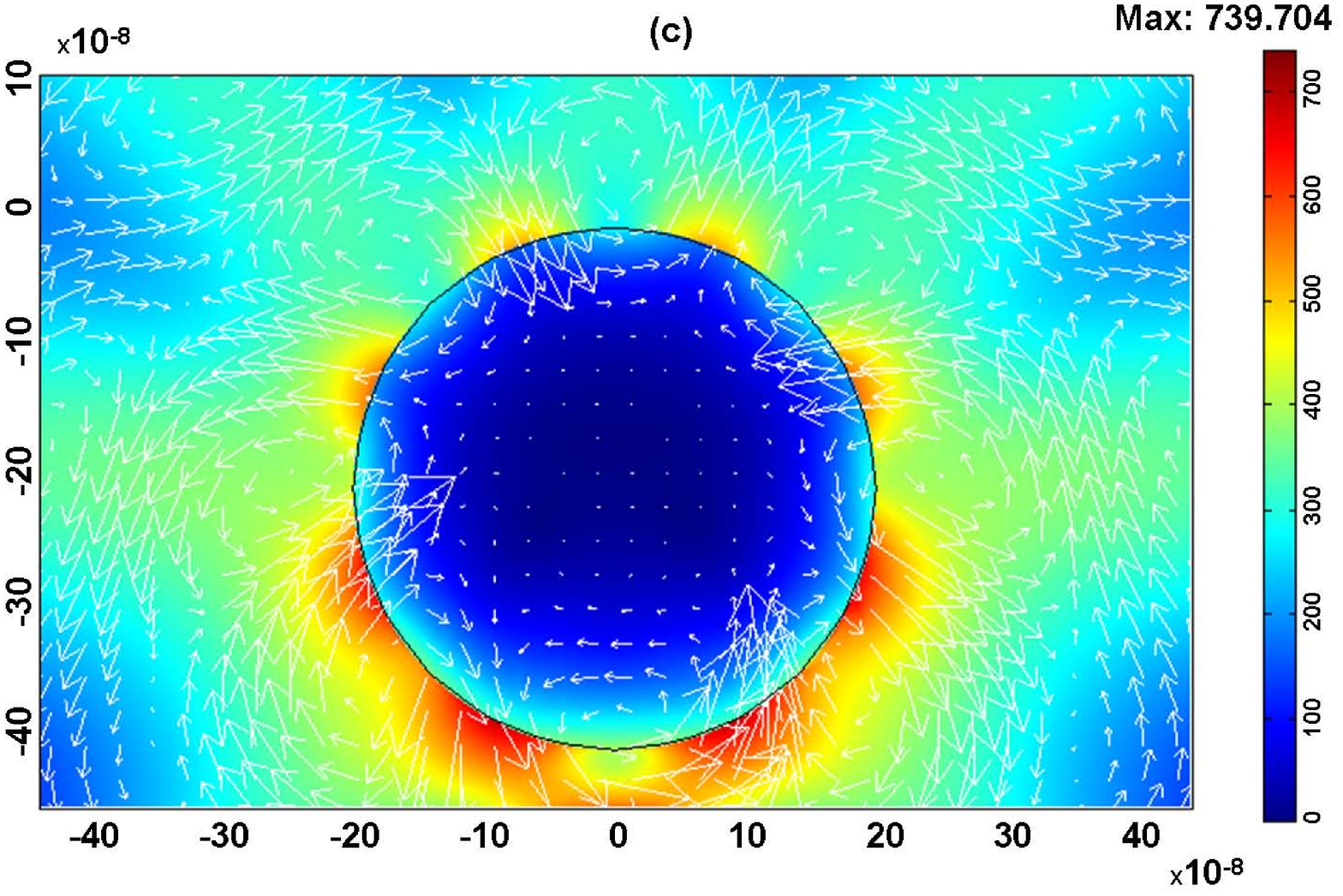}
\end{minipage}
\caption{(a) {\it Ag} cylinder (radius $r= 200nm$) illuminated by
the same beam as in Figs. 3(a) - 3(c) at $\lambda= 300.0nm$
(refractive index $n_{Ag}= 1.513 + i0.955$; $LSP_{42}$ cylinder
resonance). Magnetic field norm $|{\bf H_{z}(r)}|$ (in $A/m$ (SI
units))in colors, and time - averaged energy flow $<{\bf S(r)}>$ in
arrows (maximum arrow length $= 3067.37eV/(nm^{2}\cdot s)$). (b)
Cylinder response in time - averaged energy flow norm $|<{\bf
S(r)}>|$, averaged over the area $A= \pi(220^{2}-200^{2})nm^{2}=
26389.38nm^{2}$) of an annulus surrounding the particle (see the two
concentric circles in Fig. 4(a)) versus wavelength. The shown
plasmon resonance peaks are: $LSP_{42}$, $LSP_{32}$ and $LSP_{51}$
located at $\lambda= 300.0nm, 317.9nm$ and $357.3nm$ respectively.
(c) Detail of the electric near field ${\bf E(r)}$ (in $V/m$ (SI
units)) distribution in both norm (colors) and vector (arrows) for
the same situation as in Fig. 4(a).}
\end{figure}

\begin{figure}[htbp]
\begin{minipage}{.49\linewidth}
\centering
\includegraphics[width=7cm]{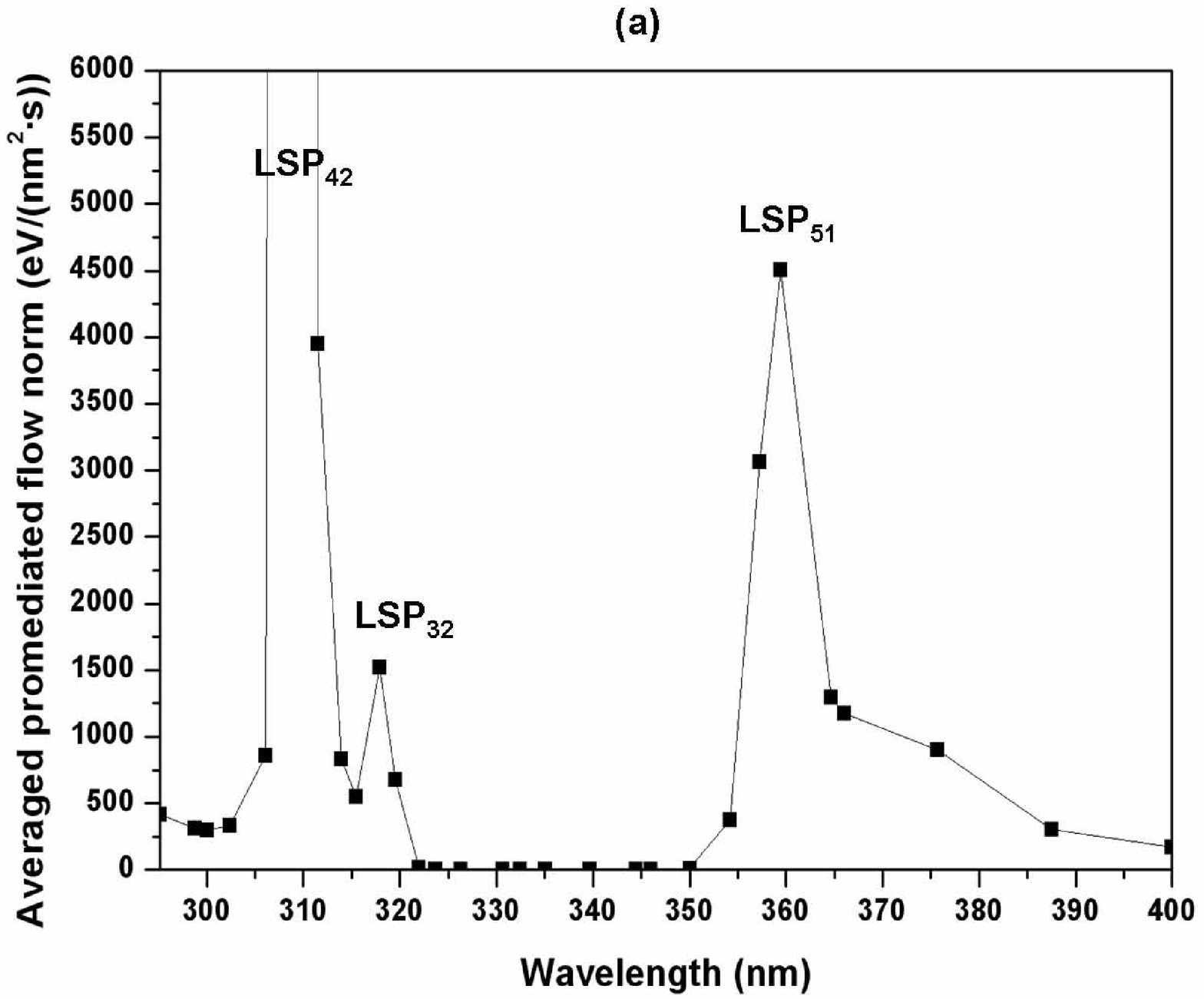}
\end{minipage}
\begin{minipage}{.49\linewidth}
\centering
\includegraphics[width=7cm]{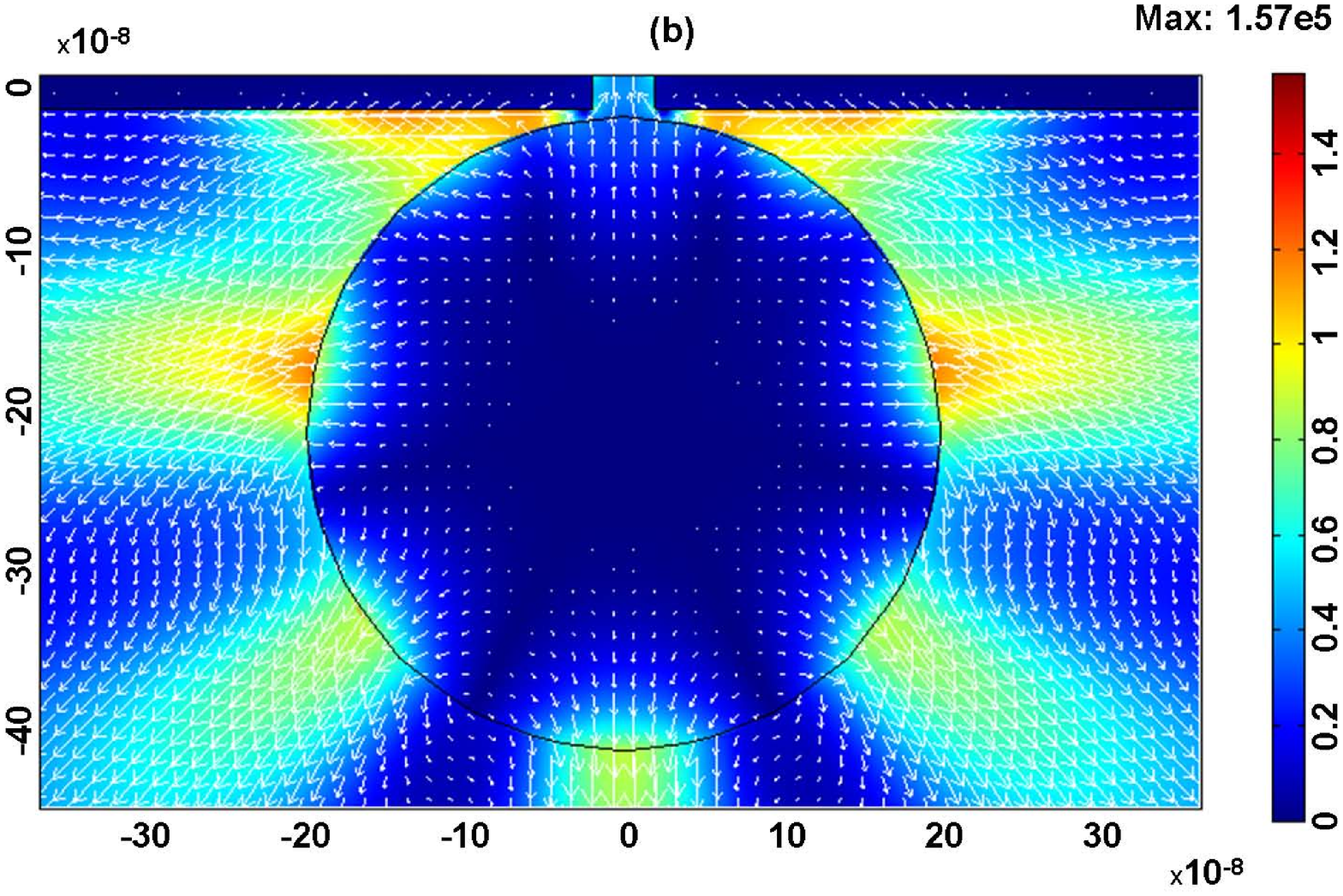}
\end{minipage}
\begin{minipage}{.51\linewidth}
\centering
\includegraphics[width=7cm]{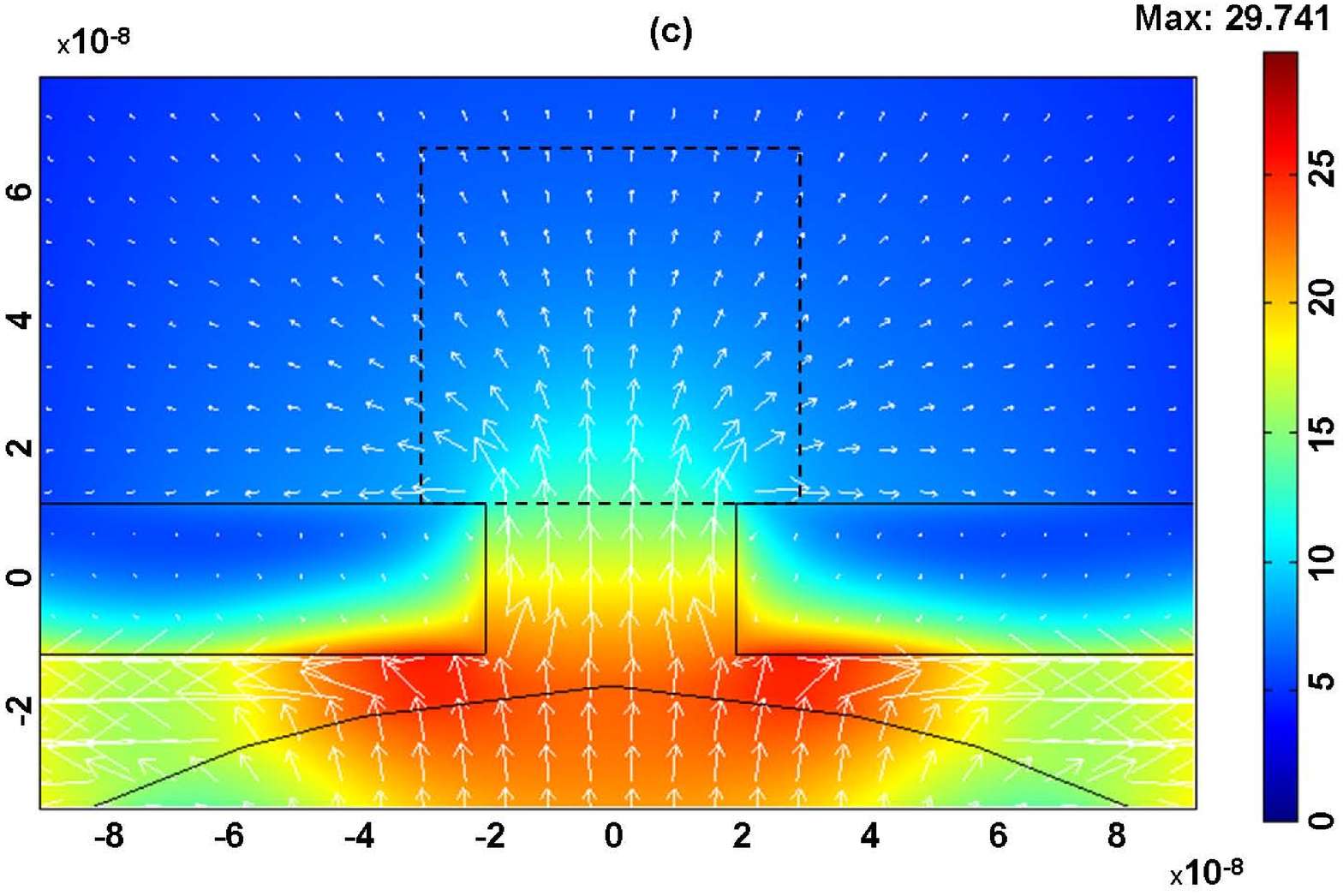}
\end{minipage}
\begin{minipage}{.51\linewidth}
\centering
\includegraphics[width=7cm]{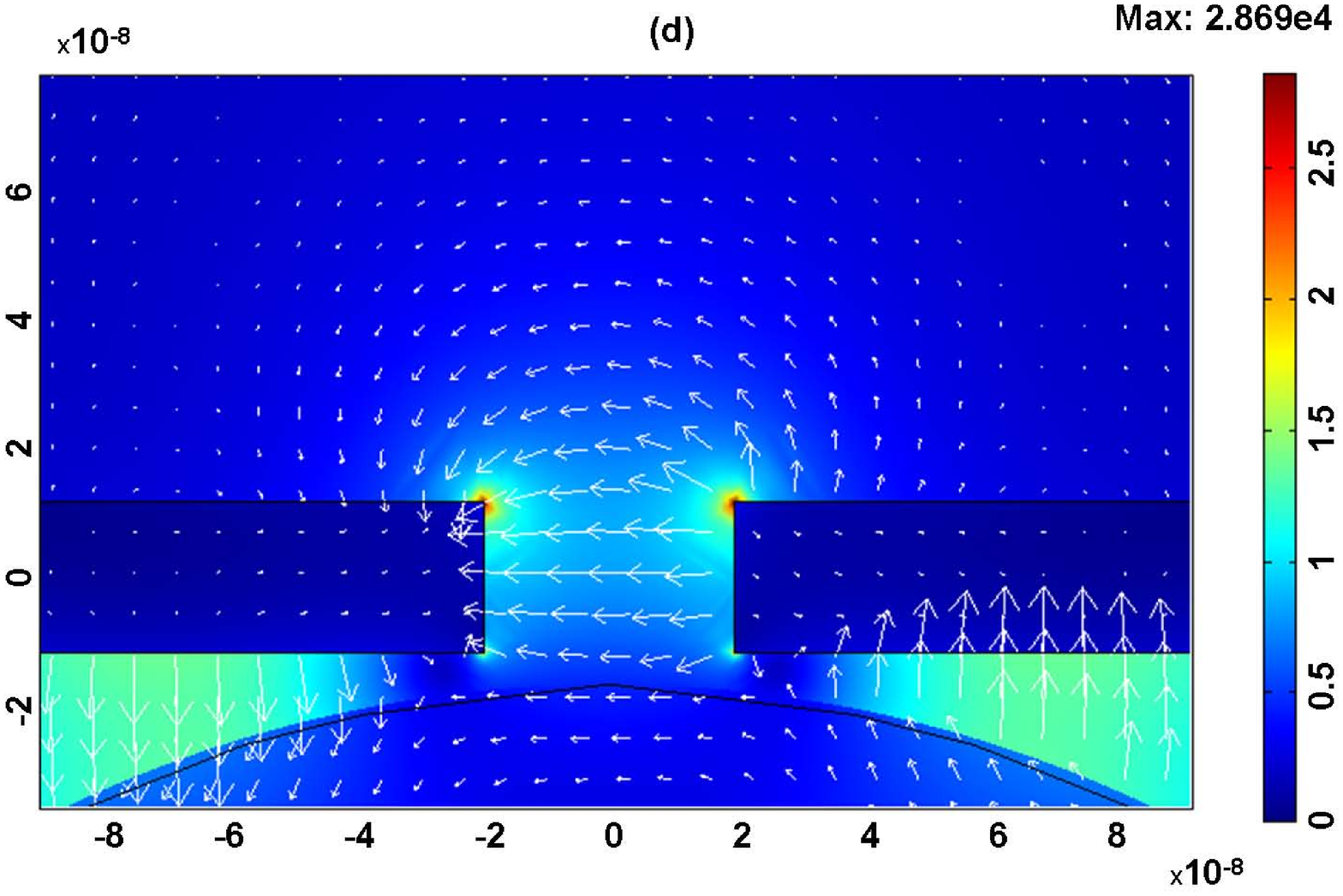}
\end{minipage}
\caption{(a) Response of the cylinder - slit/slab combination (i. e.
those in Figs. 1(a) and 4(a)) in time - averaged energy flow norm
$|<{\bf S(r)}>|$, averaged over the area of the same square monitor
as in Figs. 1(a) - 1(b) and 3(a) - 3(b) (see the square of Fig.
5(c)). The shown resonance peaks are: $LSP_{42}$ (it reaches the
value of $\approx 120000eV/(nm^{2}\cdot s)$), $LSP_{32}$ and
$LSP_{51}$ located at $\lambda= 310.0nm$, $317.9nm$ and $359.4nm$,
respectively. The illumination beam has the same $\sigma$ as in
Figs. 1(a) - 1(c). (b) Magnetic near field norm ${\bf |H_{z}(r)|}$
(in $A/m$ (SI units)) in colors, and averaged energy flow $<{\bf
S(r)}>$ in arrows on illumination as in \figurename{ 5(a)} at
$\lambda= 310.0nm$ (refractive index $n_{Al}= 0.294 + i3.74$;
$n_{Ag}= 1.323 + i0.647$; the $LSP_{42}$ cylinder resonance is
excited). This particle is placed at a distance of {\it 5nm} between
its surface and the entrance plane of the {\it Al} slab. (c) Detail
of Fig. 5(b). Magnetic field norm ${\bf |H_{z}(r)|}$ (in $A/m$ (SI
units)) in colors, and time - averaged energy flow $<{\bf S(r)}>$ in
arrows (maximum arrow length $= 979909.24eV/(nm^{2}\cdot s)$). (d)
Detail of the electric near field ${\bf E(r)}$ (in $V/m$ (SI units))
distribution in both norm (colors) and vector (arrows) for the same
configuration as in Fig. 5(c).}
\end{figure}

Since transmission enhancement through the slit is linked to the
induction of resonantly large field energies in its entrance that
couple with the subwavelength slit mode, other larger plasmonic
particles whose LSP modes produce such strong localized fields
should also give rise to the same phenomenon. However, in this case
the perturbation of the resonant wavelength introduced by the
combination aperture - particle upon that of the particle alone
should be more noticeable. Let us study it. Figure 4(a) shows the
spatial distribution of magnetic field magnitude and energy flow in
the plasmon resonance $LSP_{42}$ of an illuminated isolated {\it Ag}
cylinder of radius $r= 200nm$ at $\lambda= 300.0nm$. As shown by the
energy flow norm spectrum of this particle averaged near the
particle surface in Fig. 4(b) (i. e. the annulus of Fig. 4(a)), this
is not the plasmon excitation with the larger quality factor, which
is the one at $\lambda= 317.9nm$ (corresponding to $LSP_{32}$). A
detail of the distribution of electric field at $\lambda= 300.0nm$
is presented in Fig. 4(c). However, when this particle is placed
close to the aperture entrance, the behavior of the system
concerning slit transmission, beyond the redshift suffered by the
cylinder modes, depends on the particular LSP mode excited on the
particle surface. Two examples can be found in both the $LSP_{32}$
and $LSP_{51}$ modes, whose effect on enhancement in the slit
transmission is lower and higher, respectively, than expected from
their responses in energy concentration around the isolated cylinder
surface (compare these resonant peaks in Fig. 4(b) for the cylinder
alone with those of the transmitted energy in Fig. 5(a) when this
particle is placed at the entrance of the slit). Nevertheless, the
excitation of the red-shifted $LSP_{42}$ on the cylinder below the
slit does render, however, a huge transmission enhancement through
the slit. Namely, at $\lambda= 310.0nm$, there is a dramatic change
in the resonantly transmitted light by this combined system into the
other side of the slit, as shown in Figs. 5(a) - 5(d). There the
magnetic field, the electric field and the averaged energy flow
transmitted at $\lambda= 310.0nm$ over the aforementioned square
monitor placed at the exit of the slab, show a huge transmitted
averaged energy flow, of large quality factor, which is comparable
to that shown in Fig. 3(a) when the dipolar plasmonic cylinder was
placed instead. Also, this $LSP_{42}$ resonance wavelength in
presence of this larger cylinder is quite separated from the larger
energy resonance $LSP_{32}$ of this cylinder alone, as seen on
comparing Fig. 5(a) and Fig. 4(b). In addition, a comparison of Fig.
3(b) and Fig. 3(c) for the small cylinder placed below the aperture,
with Figs. 5(b) - 5(d) that show the same quantities for this larger
cylinder placed instead, evidences that such a large transmission
enhancement is also produced by a large but rather delocalized
energy distribution below the aperture, with local values in their
spatial distribution, similar to the former. Hence, we conclude that
from the point of view of concentrating most of the available
excited energy in an effective way through transmission by producing
high localized fields, it is better to employ small (even dipolar)
plasmonic particles.

\section{Comparison with extraordinary transmission by nanojet superfocusing}

\begin{figure}[htbp]
\begin{minipage}{.48\linewidth}
\centering
\includegraphics[width=7cm]{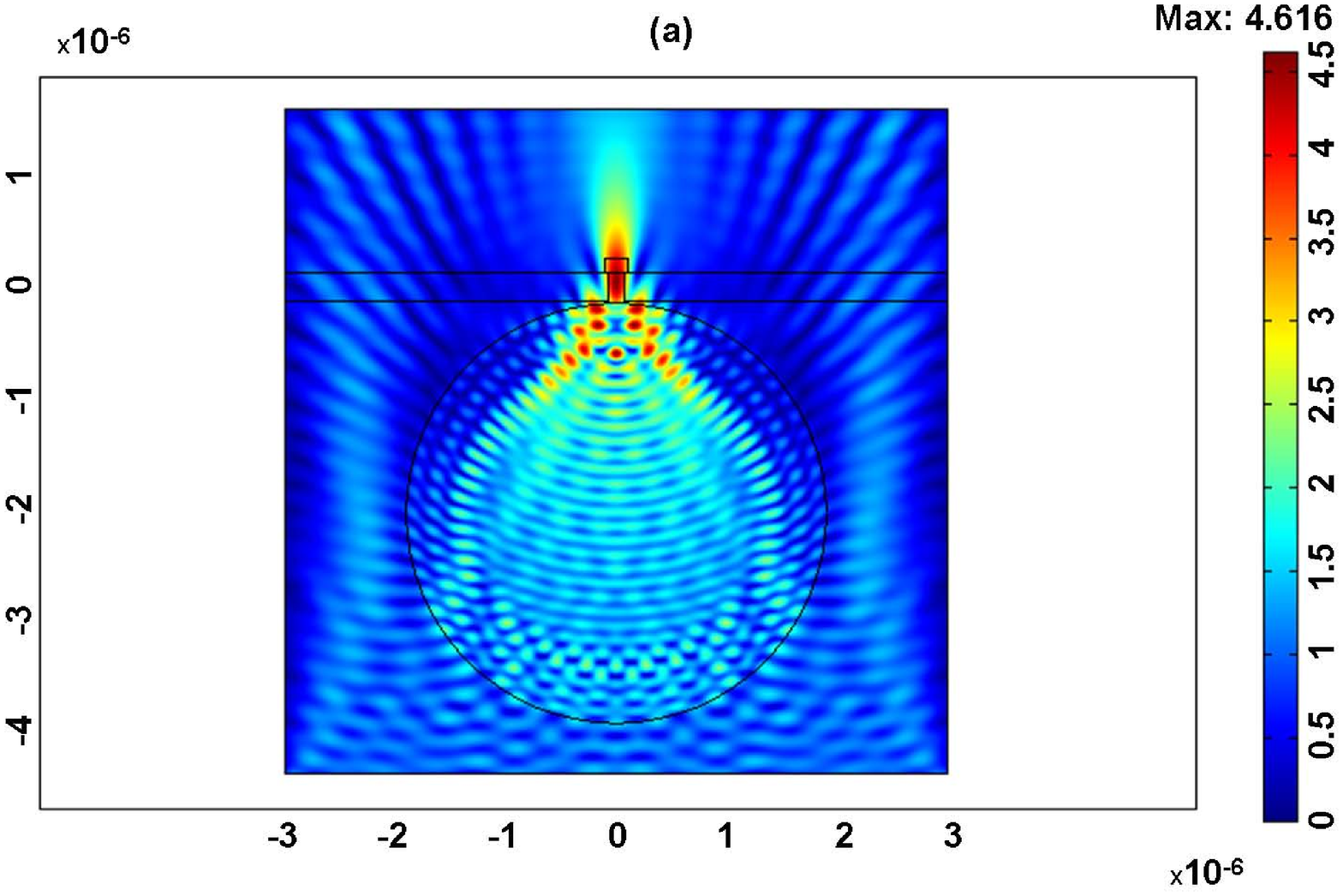}
\end{minipage}
\begin{minipage}{.48\linewidth}
\centering
\includegraphics[width=7cm]{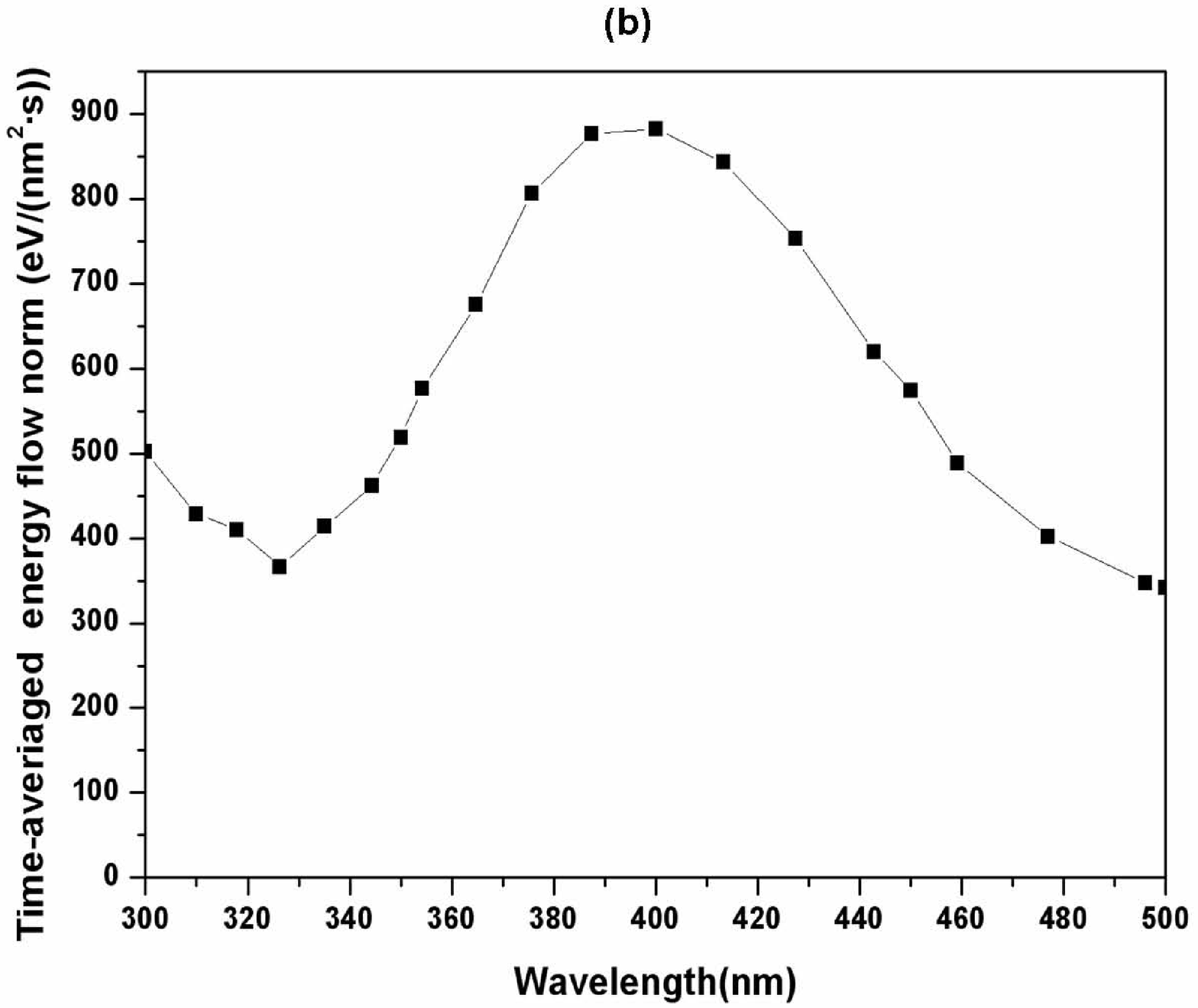}
\end{minipage}
\begin{minipage}{.48\linewidth}
\centering
\includegraphics[width=7cm]{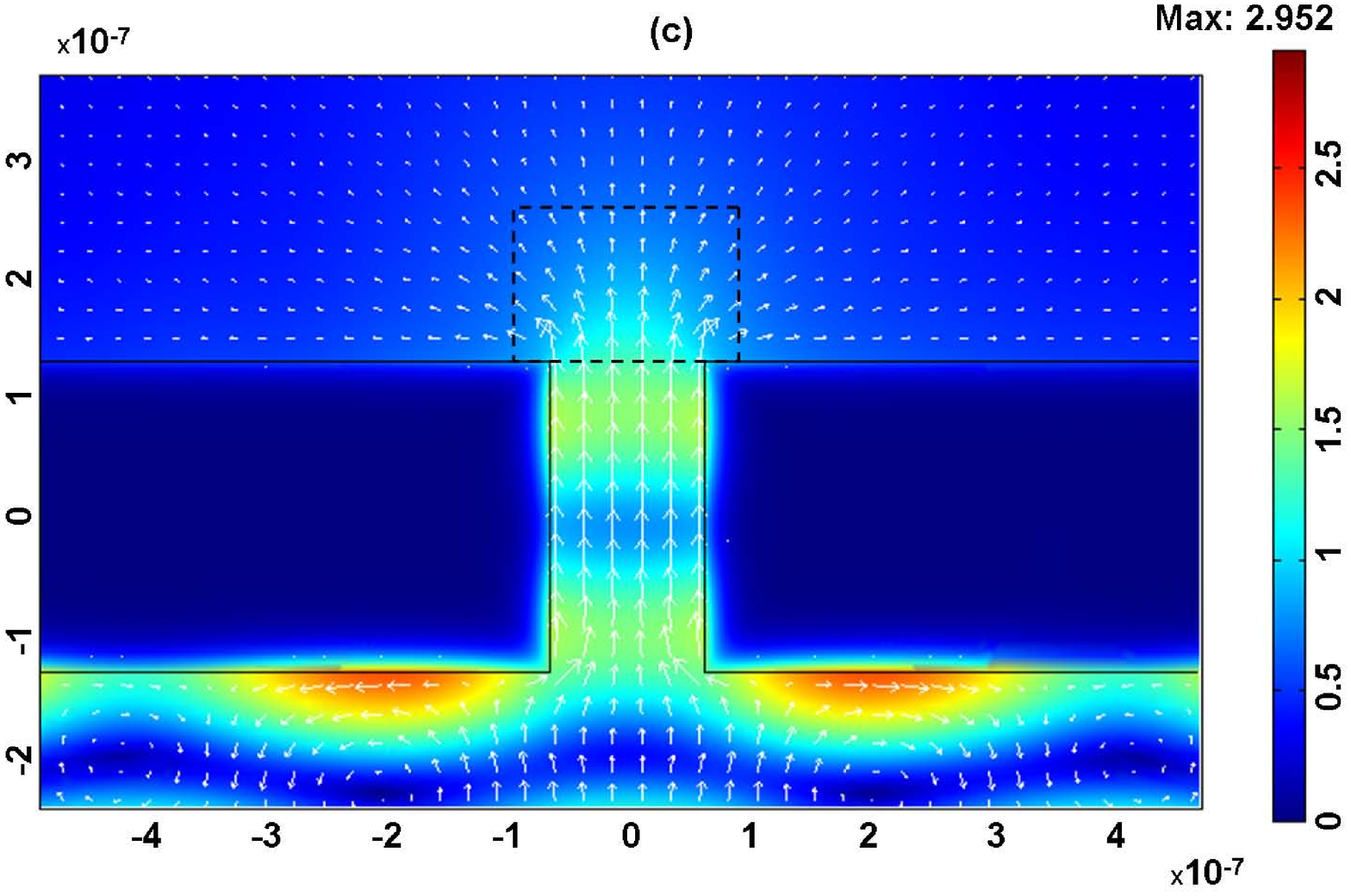}
\end{minipage}
\begin{minipage}{.48\linewidth}
\centering
\includegraphics[width=7cm]{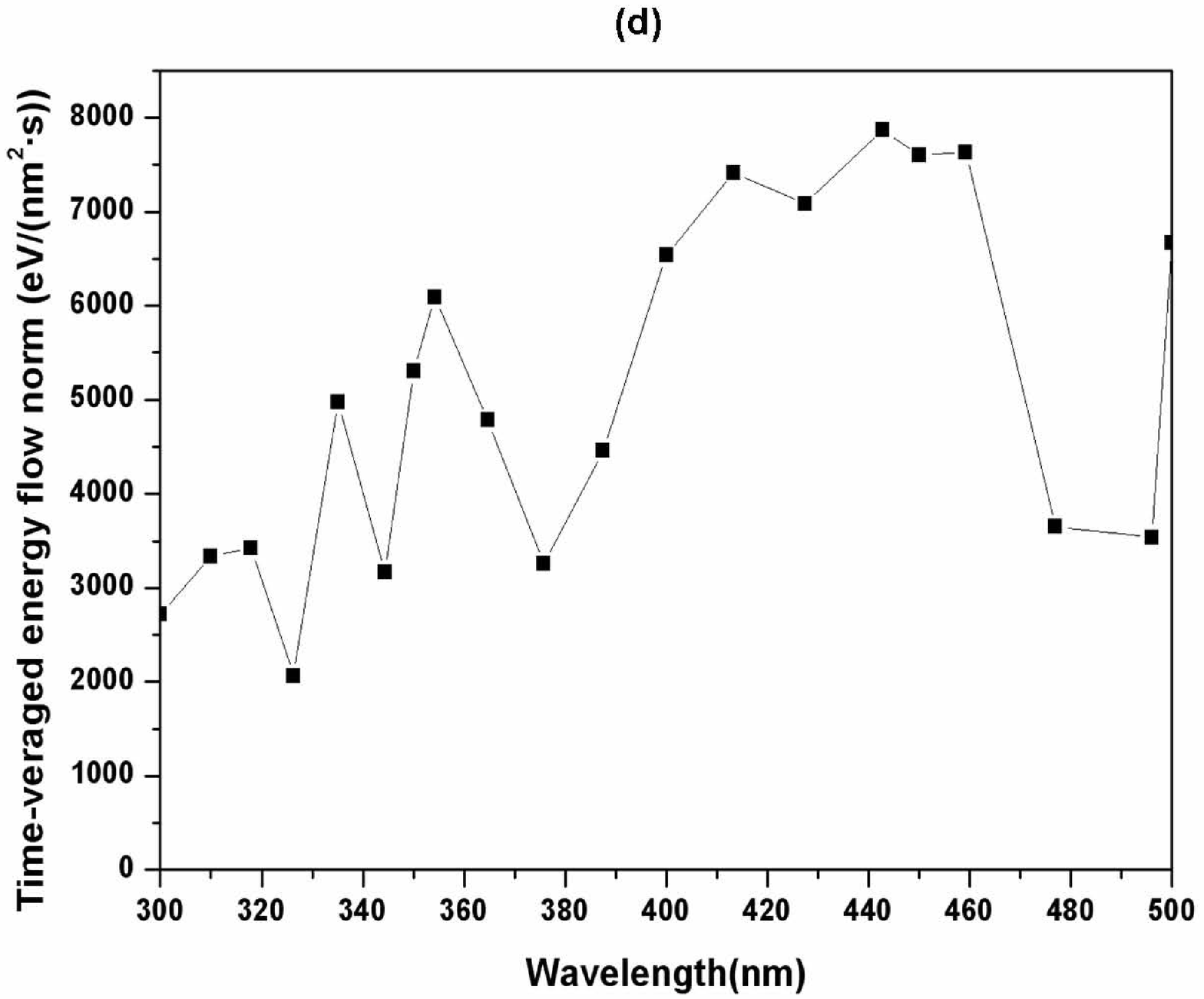}
\end{minipage}
\begin{minipage}{.98\linewidth}
\centering
\includegraphics[width=7cm]{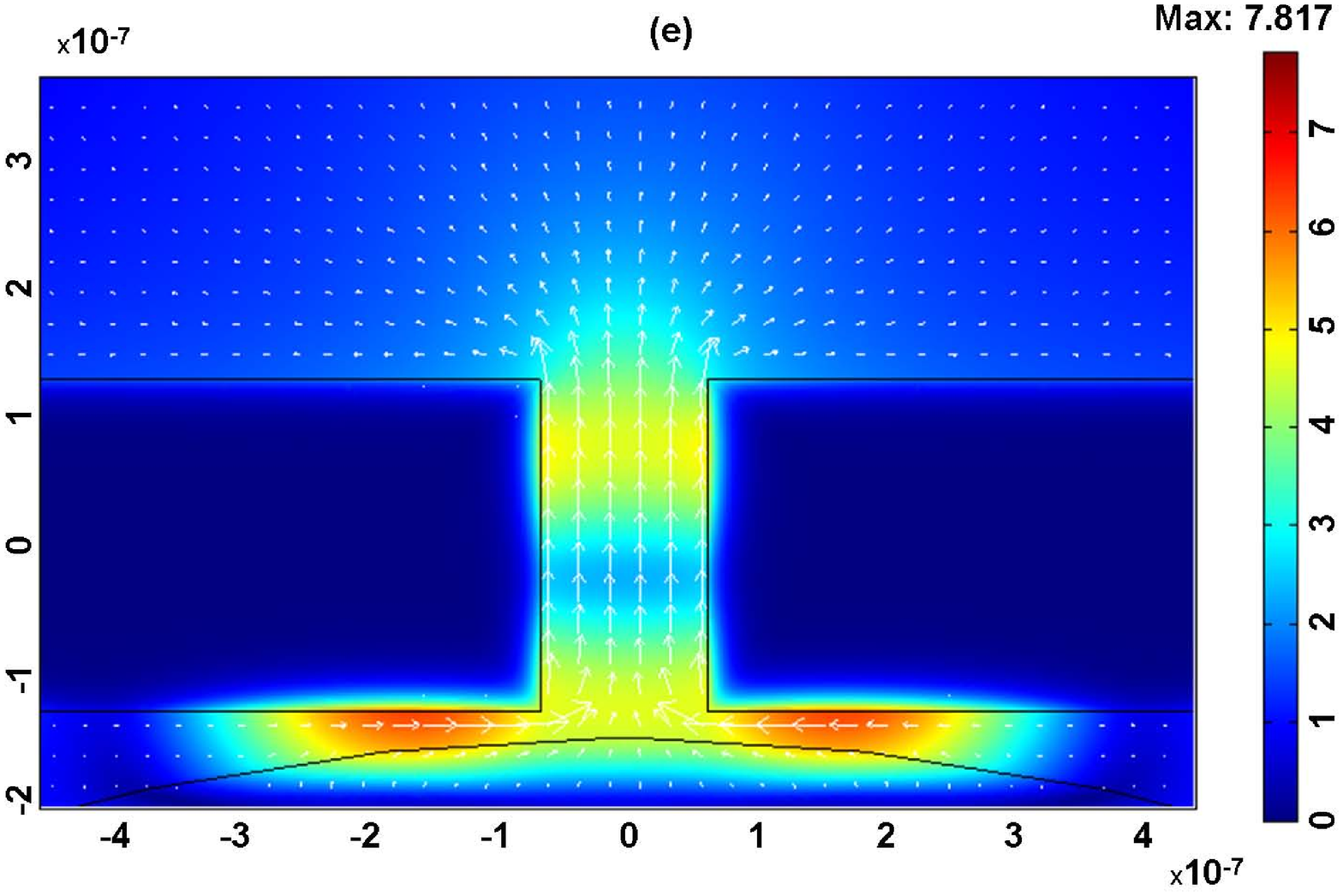}
\end{minipage}
\caption{ (a) Spatial distribution of magnetic field norm $|{\bf
H_{z}(r)}|$ (in $A/m$ (SI units)) in colors. Time - averaged energy
flow (amplitude $|<{\bf S}>|= 19364.69eV/(nm^{2}s)$ for the hottest
point of the nanojet whose width is $w\approx 120nm$) in an $\alpha
- crystalline$ $Si0_{2}$ cylinder (refractive index $n_{silica}=
1.558$; cylinder radius $r= 1.9\mu m$) illuminated from below by a p
- polarized unit amplitude rectangular beam (width $w= 6\mu m$) at
$\lambda= 400.0nm$. The slab and slit to be used later have been
drawn to see their relative positions in the following figures. (b)
Transmission of an isolated metallic {\it Al} layer (width $D= 6\mu
m$; thickness $h= 258.30nm$; slit width $d= 129.15nm$) in time -
averaged energy flow norm $|<{\bf S(r)}>|$, averaged in the
rectangular monitor of area $A= ((3/2)d)d= 25019.58nm^{2}$ at the
exit of the aperture (see rectangle of Fig. 6(c)). A
supertransmission peak arises at $\lambda= 400.0nm$ (corresponding
to the slit $TE_{10}$ mode). (c) Detail of the aperture whose
response is studied in Fig. 6(b), illuminated at $\lambda= 400.0nm$
(refractive index $n_{Al}= 0.490 + i4.86$). Magnetic field norm
$|{\bf H_{z}(r)}|$ (in $A/m$ (SI units)) in colors, and $<{\bf
S(r)}>$ in arrows (maximum arrow length $= 4829.44eV/(nm^{2}\cdot
s)$). (d) Response of the cylinder - slab combined system in time -
averaged energy flow norm $|<{\bf S(r)}>|$, averaged in the same
monitor as in Fig. 6(b). The system reaches its highest transmission
at $\lambda= 442.8nm$. (e) Detail of transmission in the aperture
evaluated as in Fig. 6(b) when the combined cylinder - slab system
is illuminated as before at $\lambda= 442.8nm$ (refractive index
$n_{Al}= 0.598 + i5.38$; $n_{silica}= 1.553$). Magnetic field norm
$|{\bf H_{z}(r)}|$ (in $A/m$ (SI units)) in colors, and $<{\bf
S(r)}>$ in arrows (maximum arrow length $= 43480.41eV/(nm^{2}\cdot
s)$). The distance between the cylinder surface and the entrance
plane of the {\it Al} slab is {\it 20nm}. }
\end{figure}

As concluded from the above calculations, enhancements and
localization of the near field entering the subwavelength aperture
at conditions such that the slit transmitting eigenmodes are
excited, is behind the phenomenon of transmission enhancement. Then,
one may inquire if any effect that gives rise to a large and highly
localized field intensity distribution in the aperture, even if it
is not resonant, will produce a similar phenomenon. To this end, we
shall next consider a situation of superfocusing into the slit. This
requires a large dielectric particle that acts as a near field lens,
(see e. g. \cite{Kino1994}). The geometrical parameters are now
adapted to the spot size of the focused intensity produced by the
particle. This superfocusing is done with a $SiO_{2}$ \cite{Palik}
cylinder of radius $1.9\mu m$, illuminated by a p-polarized
rectangular-profile beam of width $w= 6\mu m$. A {\it nanojet}
\cite{Taflove2004, Taflove2005, Taflove2006, Taflove2009} produced
by superfocusing in the cylinder, of approximate half width at half
maximum $HWHM= 120nm$, is produced at the opposite side of the
surface of illumination in the cylinder, as illustrated for this
particle alone in Fig. 6(a) for illumination at $\lambda= 400nm$, in
which we have also drawn the slab with the slit to indicate the
relative position between them and the cylinder when both objects
are subsequently present. The enhancement and concentration of
energy flow in the nanojet created on top of the cylinder is already
evident in this figure. The slit that we next consider of width
$d=129.15nm$, is again in an {\it Al} slab, this time of thickness
$h=258.3nm$ and width $D=6\mu m$. The energy flow transmitted by
this slit alone in a rectangular monitor of area $((3/2)d)d=
25019.58nm^{2}$ is plotted in Fig. 6(b) as the incident wavelength
$\lambda$ of the aforementioned beam varies. As shown, there is a
maximum of transmission of the slit at $\lambda=400nm$. In Fig. 6(c)
one sees the magnetic field magnitude and time-averaged energy flow
spatial distribution $<{\bf S(r)}>$ near the slit corresponding to
this illumination wavelength of maximum transmission by the slit
alone.

If we now place the above cylinder in front of the slit, Fig. 6(d)
shows versus $\lambda$ the transmission  of the energy flow into the
above mentioned evaluation rectangular monitor. Also in this figure
one sees that near the supertransmission wavelength, the shape of
the transmission spectrum does not appreciably change except for
ripples due to reflections in the cylinder - slit/slab cavity, even
though its peak due to the presence of the nanojet increases by a
factor of 8 and is red-shifted. As shown in the detail of Fig. 6(e),
an enhancement of the aperture supertransmission is now obtained.
This figure corresponds to the maximum of this quantity as plotted
in Fig. 6(d), which occurs at $\lambda= 442.8nm$. Hence, again, the
resonant transmittance of the aperture is red-shifted by the
presence of the particle. Since the nanojet is not a resonance, but
a subwavelength focusing effect, and the $SiO_{2}$ cylinder
refractive index does not yield as much feedback with the
aperture/slit via multiple reflections between them, the magnitude
of this enhancement is much smaller, however, than that produced by
the interaction between a particle plasmon resonance and the mode
$TE_{10}$ of the subwavelength slit, as shown in the previous
examples with {\it Ag} particles.

\section{Discussion and conclusions}
In the above calculations, we have employed a region of wavelengths
in the UV for observations of LSP excitation and in the violet for
nanojets. As mentioned in the beginning of Section 3, one may
equally choose other parameters and materials to observe similar
phenomena. For instance, concerning the excitation of LSPs we have
done other computations with different wavelengths, sizes and
materials for the slab and the particle, obtaining similar
qualitative results. For example, if one uses an {\it Au} particle
of radius $a= 50nm$, this cylinder alone presents a $LSP_{11}$
resonance peak at $\lambda= 496nm$ of time-averaged energy flow of
about $476eV/(nm^2 \cdot s)$, with lower quality factor than that of
Fig. 2(a), and whose line shape is rather similar to that of an {\it
Au} sphere of radius $a= 20nm$, (this latter shown in Fig. 12.17 of
\cite{Bohren}). Then by employing an {\it Al} slab with a slit whose
parameters $D$, $h$ and $d$ are scaled with respect to those of
Section 3 by a factor $5/3$, one obtains in the range of wavelengths
around $\lambda= 496nm$ a tail of the transmitted energy for the
slit alone, (with a value of about $133eV/(nm^2 \cdot s)$ near that
wavelength), and a peak of transmission at $\lambda=501nm$ of
$450eV/(nm^2 \cdot s)$ for the combination slit-cylinder. This
amount of transmitted energy is already similar to that around the
cylinder alone, but it is more than 3 times larger than that of the
slit alone. Analogous results are obtained concerning nanojet
excitation with other materials and sizes.

Thus we conclude by stating the universality of the enhancement of
transmission through a subwavelength aperture by excitation of
particle LSP resonances, or in general by reinforcing and localizing
the incident energy at the entrance of the aperture by other
procedures like e.g via a nanojet or with other means of
superfocusing. This enhancement is independent of whether or not the
aperture alone produces extraordinary transmission, providing of
course that the illumination is chosen such that a homogeneous (i.
e. propagating) eigenmode is excited in its cavity; in particular,
incident p-polarization is necessary when dealing with 2D slits.
Transmission is dominated by either the LSP or by the creation of
the nanojet. And it is is more efficient the larger their lineshape
Q factor is, depending on their sizes and hence on the resonant
wavelength perturbation of the combined system slit - particle.

We hope that these results stimulate experiments in both 2D and 3D.
In particular, if a plasmonic nanoparticle chain forms a signal and
energy transporting waveguide, the aperture may then constitute an
interesting coupling device, to which high directionality of the
transmitted light may be added if one, for instance, introduces
periodic corrugation in the slab. A similar effect may occur for
superfocused light and for nanoantennae in front of slits.

\section*{Acknowledgements}

Work supported by the Spanish MEC through FIS2009-13430-C02-C01 and
Consolider NanoLight (CSD2007-00046) research grants, FJVV is
supported by the last grant.

\end{document}